\documentclass[final]{aa}
\usepackage{xcolor}
\usepackage{graphicx}
\usepackage{txfonts}
\usepackage{lipsum}
\usepackage{subcaption}         
\usepackage{lscape}             
\usepackage{placeins}           
\usepackage{url}

\begin{document}
\nolinenumbers
   \title{Global evolution of electric fields during planet encircling dust storms on Mars }

\author{Ina Taxis\inst{1}\thanks{Corresponding author: ina.taxis@campus.lmu.de}
        \and Leonardos Gkouvelis\inst{2}\thanks{Corresponding author: gkouvelis@iaa.es}
        \and Richard A. Urata\inst{3}
        \and Melinda A. Kahre\inst{3}
        \and Amanda S. Brecht\inst{3}
}

    \institute{
   $^1$Ludwig Maximilian University, Faculty of Physics, University Observatory,
Scheinerstrasse 1, Munich D-81679, Germany\\
$^2$Instituto de Astrofísica de Andalucía (IAA-CSIC), 
Glorieta de la Astronomía s/n, E-18008 Granada, Spain\\
   $^3$Ames Research Center, Space Science Division, National Aeronautics and Space Administration (NASA), Moffett Field, CA, USA\\}

   \date{Received , 20XX}

  \abstract{Planet-encircling dust storms fundamentally reshape Martian weather and the near-surface electrostatic environment, yet the evolution of electric fields on global scales has not been previously investigated with physically consistent triboelectric modeling. We investigate the generation and evolution of electric fields during global dust storms (GDSs) using bimodal dust size distributions from the NASA Ames Mars Global Climate Model (GCM), coupled with a triboelectric charging and electrostatic diagnostic scheme that links collisional charging to the local dynamical state of the atmosphere. Focusing on the dust-lifting and buildup phase and its subsequent evolution, we quantify the electric-field energy density and discharge characteristics, including onset thresholds, event frequency, and spatial clustering. The simulations reveal broad storm-active belts of enhanced electrification, with the most frequent threshold exceedances occurring in southern low-to-mid latitudes and secondary activity in northern low-to-mid latitudes. Modeled near-surface electric fields reach $10^{2}$--$10^{3}\ \mathrm{V\,m^{-1}}$, with domain-maximum values of several $10^{4} \ \mathrm{V\,m^{-1}}$, comparable to values inferred for smaller-scale dust phenomena. The results indicate that electric-field generation is controlled by the interplay between dust loading, turbulence-driven collisional activity, and conductivity-dependent charge relaxation, with diurnal conductivity variations strongly suppressing daytime electric-field buildup and most events remaining in the weak glow or Townsend discharge regime. While the model captures the large-scale distribution of electrically favorable conditions, the predicted spatial extent of activity likely represents an upper bound, as small-scale turbulent structures are not fully resolved. These results provide a quantitative framework to identify regions where electrostatic discharges are most likely during GDSs and to inform instrument design, power-system protection, and operations planning for future robotic and human missions.}

\maketitle

\section{Introduction}\label{sec: 1}

Electric fields are well-known phenomena that manifest differently across planetary environments. On Earth, they are most prominent during thunderstorms in the troposphere, where magnitudes can reach several kilovolts per meter (e.g.~\citealt{Adzhieva2020JPhCS1604a2016A}). Analogously, dust particles in other planetary atmospheres can acquire charge through triboelectric processes when present at sufficiently high number densities and collisional rates, such as during dust storm events (\citealt{Abdelaal2025SoSyR..59...71A}).

A large fraction of the Martian surface is covered by fine dust that is frequently lifted into the atmosphere and plays a key role in both thermal and dynamical processes, from the lower atmosphere up to the thermosphere (e.g.~\citet{Sheel2016}; \citet{Gkouvelis2020, Gkouvelis2020b,Aoki2022, Soret2022, Soret2025}). Increased concentrations of airborne dust undergo triboelectric, or contact, electrification (\citealt{Izvekova2020JPhCS1556a2071I}). While the detailed mechanisms of charge transfer remain under investigation, proposed pathways include the exchange of electrons, ions, chemical species, or particulates. Laboratory and field studies have quantified the charge acquired by direct contact, friction, and collisions, showing a dependence on particle-size contrast and material composition. In general, the smaller of two colliding particles tends to acquire a net negative charge, while the larger particle gains an equivalent positive charge (\citealt{Forward2009GeoRL..3613201F}; \citealt{Farrell2006}).

During dust storm events, smaller particles are entrained into atmospheric currents and dominate the upper layers, whereas larger particles settle more rapidly and remain concentrated near the surface (\citealt{Barth2016Icar..268..253B}). This vertical separation between negatively and positively charged populations leads to the buildup of electrostatic fields. The development of such fields in dust-laden vortices has been investigated under both terrestrial and Martian conditions (\citealt{Zhang2020NatCo..11.5072Z}; \citealt{Ruf2009GeoRL..3613202R}). On Earth, dust devils and sandstorms typically generate electric fields of up to a few tens of kilovolts per meter (\citealt{HarrisonEtAl2017}). On Mars, model results and laboratory experiments suggest fields of several hundred volts per meter up to a few kilovolts per meter. Small- to medium-scale vortices, commonly referred to as dust devils, have been the focus of numerous studies predicting field strengths approaching breakdown thresholds (e.g.~\citealt{Farell2003}; \citealt{Farrell2006}; \citealt{Delroy2011}; \citealt{Franzese2018}; \citealt{Melnik1998}; \citealt{Sheel2025}). The field strength is controlled by vortex diameter, particle velocity differentials, and atmospheric conductivity. Although direct \textit{in-situ} measurements are sparse, existing observations are consistent with model predictions (\citealt{Franzese2018}; \citealt{Delroy2011}; \citealt{Ruf2009GeoRL..3613202R}).  

Recent in situ observations of electrical discharges on Mars suggest that triboelectric activity is not directly controlled by bulk dust abundance, but is instead associated with turbulent and convective structures such as dust devils and storm fronts \citep{Chide2025}.  Motivated by this, we adopt a modeling framework in which collisional charging is explicitly linked to the local dynamical state of the atmosphere, rather than prescribed through uniform conditions.

The magnitude, spatial organization, and discharge regime of electrical activity on Mars are therefore of considerable interest, particularly in light of the recent detections of triboelectric discharges in dusty convective environments. \citet{Bertrand2020} investigated dust-storm evolution, including the time-dependent lifting and settling of particles, while \citet{Kok2006GeoRL..3319S10K} noted that electric forces may influence particle motion, although their magnitude remains uncertain. In addition, laboratory studies under Mars-like conditions (e.g.~\citealt{Kim2013JAChS.135.4910K}; \citealt{Navarro2010JGRE..11512010N}) indicate that chloride salts exposed to plasma or ultraviolet radiation in oxidizing environments can form chlorates and perchlorates, suggesting possible chemical consequences of electrification.

Unlike terrestrial dust storms, Martian dust events can evolve to global scale, obscuring surface features for weeks, with visible optical depths exceeding $\tau \approx 5$ (\citealt{Bertrand2020}; \citealt{Haberle2019}). Global dust storms (GDSs) lack strict periodicity but typically occur during northern autumn and winter at intervals of roughly 3--5~years (\citealt{Smith2019}). Their onset and evolution depend on a complex interplay among solar forcing, atmospheric circulation, surface wind thresholds, and dust-reservoir availability.

In this work, we focus on the 2018 global dust storm, one of the most recent and well-characterized events both observationally and through modeling. The evolution of the 2018~GDS is simulated using the NASA Ames Mars Global Climate Model (GCM; \citealt{Urata2024LPICo3007.3337U}; \citealt{Urata2025Icar..42916446U}; \citealt{Kahre2023}). We couple the model’s bimodal dust distribution to a triboelectric charging parameterization and a quasi-electrostatic field solver to estimate the spatiotemporal structure and intensity of electric fields. This paper presents the construction and implementation of the charging and field-solving framework, quantifies field magnitudes and energy densities, assesses discharge regimes and likelihood, and discusses implications for future observations and mission design. Section~\ref{sec:2} describes the NASA Ames Mars GCM, the dust-charging scheme, and the quasi-static electric-field model. Section~\ref{sec:3} presents the results, followed by discussion and conclusions in Section~\ref{sec:4}. A detailed description of the numerical tools developed for this work is provided in Appendices~\ref{app1}-\ref{app3}.

\section{Simulations } \label{sec:2}

\subsection{NASA Ames Mars Global Climate Model}

The NASA Ames Mars Global Climate Model (GCM; \citealt{Bertrand2020, Urata2025Icar..42916446U}) is a state-of-the-art model that builds upon the legacy NASA Ames GCM (e.g.~\citealt{Haberle2019, Kahre2023}). Its physical parameterizations include a Mellor--Yamada level~2 boundary-layer turbulence scheme and a two-stream correlated-$k$ radiative transfer solver. The radiative transfer scheme computes heating rates from gaseous $\mathrm{CO_2}$ and $\mathrm{H_2O}$ and from suspended dust. 

Surface topography is based on Mars Orbiter Laser Altimeter (MOLA) data, averaged to the model’s horizontal resolution. Surface albedo and thermal inertia are derived from Viking and MGS/TES observations. The surface and subsurface temperatures are predicted using a semi-implicit heat-transfer scheme that balances radiative, sensible, latent, and conductive fluxes at the upper boundary, with the lower boundary assuming zero net heat flux. 

The dynamical core is the NOAA/GFDL Finite-Volume Cubed-Sphere (FV3) solver (\citealt{Putman2007JCoPh.227...55P, harris2021fv3}), which maps the planet onto six cube faces subsequently projected onto a sphere. The simulations presented here employ a horizontal resolution of $48\times48$ grid points per cube face, corresponding to an effective grid spacing of approximately 110~km. The vertical grid comprises 36 hybrid-sigma pressure levels, with a model top at $0.002~\mathrm{Pa}$.

The dust cycle is driven by a daily-mean global dust climatology map constructed to reproduce the Mars Year~34 annual cycle, which includes the 2018 Global Dust Storm (GDS; \citealt{Montabone2020JGRE..12506111M}). Dust is lifted when the modeled column optical depth is lower than the corresponding value in the prescribed climatology. Once lifted, dust is advected by the model’s resolved circulation.
Here, dust is represented as two independent log-normal populations, via 'small' and 'large' modes, transported separately (bimodal).
The globally averaged dust cycle obtained with the bimodal scheme reproduces the target climatology when dust coagulation effects are included (cf.~Fig.~2 in \citealt{Urata2025Icar..42916446U}).

The GCM tracks dust using a two-moment approach, in which each dust constituent is represented by prognostic number and mass mixing ratios (\citealt{Haberle2019}). Dust is assumed to follow a log-normal size distribution defined by a median radius and geometric standard deviation. While the initial (lifted) particle size is prescribed, particle growth and removal are computed prognostically. Sedimentation and coagulation lead to larger particles near the surface and smaller particles aloft.  

The bimodal dust implementation consists of two independent tracers, each with distinct log-normal distributions. These two modes interact via coagulation and exchange of mass but otherwise evolve independently. The simulation analyzed here corresponds to the 20\% small-mode bimodal configuration of \citet{Urata2025Icar..42916446U}. In this setup, 20\% (by optical depth) of the lifted dust is assigned to the small mode and 80\% to the large mode. This configuration provides the best match to Mars Climate Sounder (MCS) temperature observations, yielding the smallest atmospheric temperature bias relative to the data.

\begin{figure}[!htbp]
    \centering
    \includegraphics[width=1\linewidth]{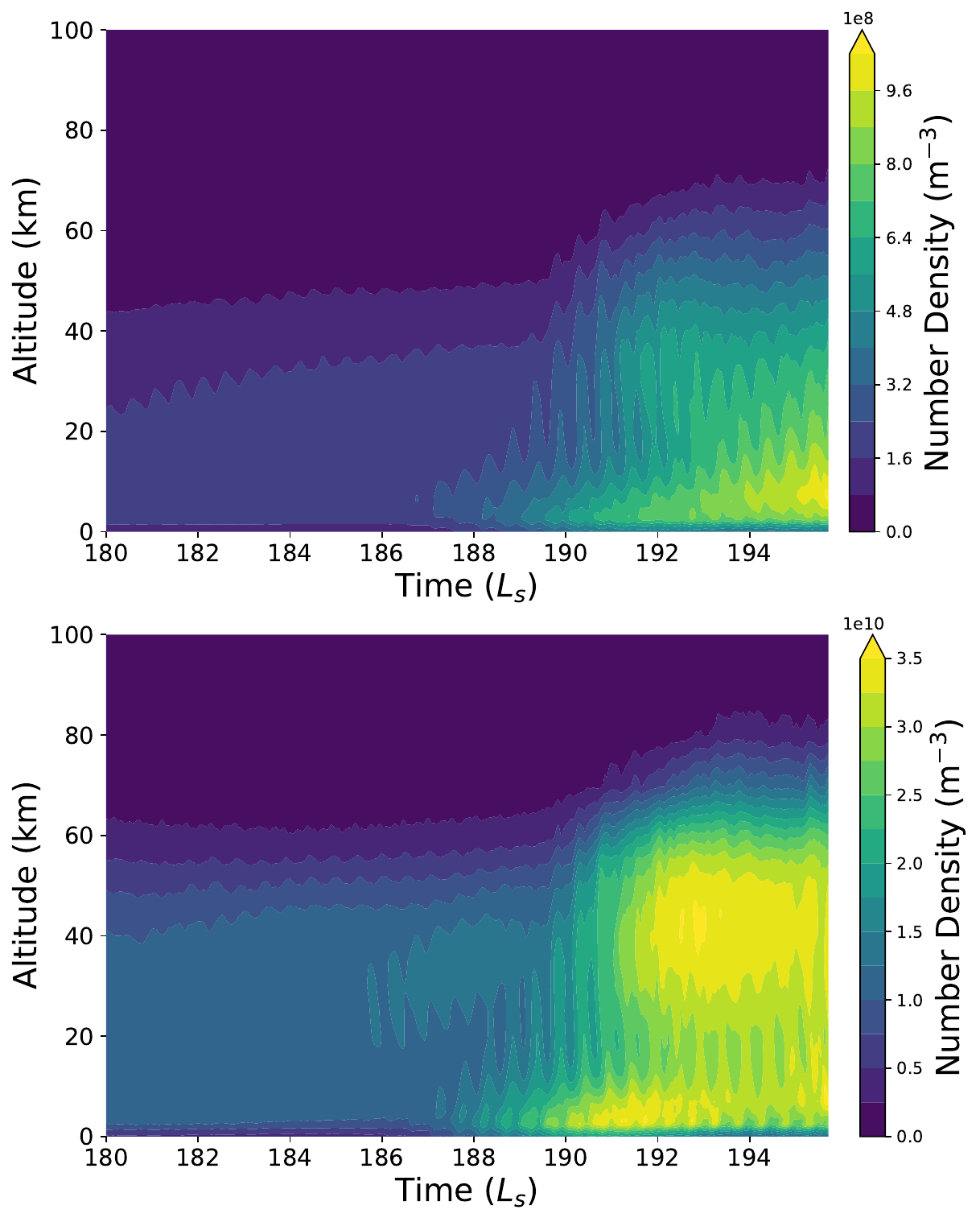}
    \caption{Number densities of small and large particles, averaged over longitude and latitude, capturing the onset of the Global Dust Storm (GDS) in Mars Year 34 (MY34) at intervals of 15 minutes.}
    \label{fig:numberdensity distribution}
\end{figure}

\subsection{Dust charging}

Charge accumulation in thunderstorms, dust devils, and other dusty environments occurs through triboelectric charging. During collisions, particles come into contact for finite times and separate with equal but opposite surface charges. The magnitude of exchanged charge depends on particle properties such as electrical conductivity, permittivity, surface microstructure, and chemical composition (\citealt{Barth2016Icar..268..253B}). 

Following \citet{Farrell1999, Desch2000, HarrisonEtAl2017}, the charge transferred during a collision between two particles can be written in terms of their mutual capacitances and the triboelectric contact potential difference.  We adopt the notation of \citet{Desch2000}: particle~1 is the larger particle (radius $r_l$) and particle~2 the smaller one (radius $r_s$). The change in the charge of particle~1 is

\begin{align}
    q_1' - q_1 = \Delta q - (1-f)\,q_{\mathrm{tot}},
    \label{eq:charge_potential}
\end{align}

where $q_1'$ (C) and $q_1$ (C) are the charges of particle~1 after and before the collision, $q_{\mathrm{tot}} = q_1 + q_2$ (C) is the conserved total charge and $\Delta q$ (C) is the triboelectric charge created by the contact potential difference $\Delta\Phi$ (V).

\begin{align}
    f = \frac{c_{11}+c_{12}}{c_{11}+c_{12}+c_{21}+c_{22}}
    \label{eq:mutual_capacitance}
\end{align}

is the charge–partition factor (dimensionless). The triboelectric term is

\begin{align}
    \Delta q = \frac{c_{12}c_{21}-c_{11}c_{22}} {c_{11}+c_{12}+c_{21}+c_{22}}\, \Delta\Phi ,
\end{align}

where $c_{ij}$ (F) are the mutual capacitances of the two-sphere system. The factor $(1 - f)\,q_{\mathrm{tot}}$ represents the fraction of any pre-existing charge that would be carried away by the small particle in the absence of a contact potential difference.

In the limit $r_l \gg r_s$, the self-capacitance of the large particle dominates the total, so $f \to 1$ and the redistribution term $(1-f)\,q_{\mathrm{tot}}$ becomes negligible. The remaining term $\Delta q$ scales with the reduced radius

\begin{align}
    r_{red} = \left( \frac{1}{r_s} + \frac{1}{r_l} \right)^{-1} \simeq r_s ,
\end{align}

because the capacitances depend on geometry through $r_{red}$. Using the dielectric properties of Martian silicate materials \citep{Farrell1999, Farrell2006, Desch2000}, one obtains

\begin{align}
    q_1' - q_1 \simeq \Delta q \approx 2668 \left( \frac{\Delta\Phi}{2~\mathrm{V}} \right) \left( \frac{r_{red}}{0.5~\mu\mathrm{m}} \right) e ,
\end{align}

where $e = 1.602\times 10^{-19}~\mathrm{C}$ is the elementary charge.

Laboratory experiments under simulated Martian conditions suggest that for silicate particles representative of Martian dust, the potential difference is approximately $\Delta \Phi \approx 2~\mathrm{V}$ (\citealt{Farrell2006, Harrison2016SSRv..203..299H, Forward2009GeoRL..3613201F}). 

The cumulative charge growth on large grains can then be estimated as

\begin{align}
Q_l' = \nu_c \, \Delta q ,
\label{eq:tribo_charging_eq}
\end{align}

where $\nu_c = \pi r_l^2 \, \Delta v \, n_s$ ($\mathrm{s^{-1}}$) is the collisional frequency (\citealt{Sheel2025}), $\Delta v$ ($\mathrm{m\ s^{-1}}$) is the interparticle relative velocity, $n_s$ ($\mathrm{m^{-3}}$) is the number density of small grains and $Q_l'$ ($\mathrm{Cs^{-1}}$) is the charge production rate.

\subsection{Interparticle relative velocity}
In previous studies, the interparticle relative velocity $\Delta v$
is typically prescribed as a constant parameter, implicitly assuming
uniform collisional conditions throughout the dust field. This
approximation does not capture the strong spatial variability of
turbulence and convection in the Martian atmosphere. To account for this, we derive $\Delta v$ directly from the resolved
dynamical state of the GCM using the diagnosed vertical eddy
diffusivity $K_{zz}$ as a proxy for local turbulent intensity.

We adopt a mixing-length closure of the form
\begin{equation}
K_{zz} \sim v_{\mathrm{turb}}\,L,
\end{equation}
which defines a characteristic turbulent velocity 
\begin{equation}
v_{\mathrm{turb}}(z) = \frac{K_{zz}(z)}{L(z)}.
\end{equation}

The interparticle relative velocity entering the collisional charging
rate (Eq. \ref{eq:tribo_charging_eq}) is then identified with this turbulent scale, 
\begin{equation}
\Delta v(z) \approx v_{\mathrm{turb}}(z).
\end{equation} 

\paragraph{Mixing length.}
The mixing length is computed using a Blackadar-type formulation,
\begin{equation}
\frac{1}{L(z)} = \frac{1}{\kappa z} + \frac{1}{\alpha z_i},
\end{equation}
where $\kappa = 0.4$ is the von Kármán constant, $z$ is altitude,
and $z_i$ is the boundary-layer depth. This formulation ensures the
correct asymptotic behavior,
\begin{align}
L &\approx \kappa z \quad \text{near the surface}, \\
L &\approx \alpha z_i \quad \text{near the top of the boundary layer}.
\end{align}

\paragraph{Boundary-layer depth.}
The boundary-layer height $z_i$ is diagnosed directly from the GCM
profiles of $K_{zz}$. For each atmospheric column, we define a threshold 
\begin{equation}
K_{\mathrm{thr}} = f\,K_{\max},
\end{equation}
with $f=0.05$ and $K_{\max}$ the maximum eddy diffusivity in the lower
atmosphere. The boundary-layer top is then defined as the lowest altitude
at which $K_{zz}(z) < K_{\mathrm{thr}}$ and remains below this threshold
at higher levels.

\paragraph{Numerical bounds.}
To avoid unphysical values, the mixing length is bounded as 
\begin{equation}
L_{\min} = \Delta z, \quad L_{\max} = 0.2\,z_i,
\end{equation}
and enforced via
\begin{equation}
L \leftarrow \min\!\left(\max(L, L_{\min}),\, L_{\max}\right).
\end{equation}

This formulation links the collisional charging rate directly to the
resolved turbulent intensity of the atmosphere. Regions of strong
convective mixing (large $K_{zz}$) produce larger relative velocities
and enhanced charging, whereas dynamically stable regions suppress
collisional activity. This avoids the assumption of spatially uniform
collision velocities and allows charging efficiency to respond
self-consistently to the evolving storm dynamics. Typical values of $\Delta v$ in storm-active regions range from $\sim$0.1 to a few m\,s$^{-1}$, depending on altitude and local
convective intensity.

\subsection{Atmospheric conductivity}

Equation~(\ref{eq:tribo_charging_eq}) predicts an unbounded linear increase of charge with time; however, atmospheric conductivity limits this growth. Charges relax until a steady state is reached, with conductivity decreasing by up to two orders of magnitude under dusty conditions. This enhances the likelihood of significant charge accumulation and even electrical breakdown. Accounting for this relaxation yields a modified balance equation:

\begin{align}
Q'_{l,s} + Q_{l,s} \left( \frac{\sigma}{\epsilon_0} \right) = \Delta q_{l,s} \, f_c ,
\label{eq:tribo_charging_with_loss}
\end{align}

where $\sigma$ ($\mathrm{S\ m}^{-1}$) is the atmospheric conductivity, $\epsilon_0$ ($\mathrm{F\ m}^{-1}$) is the permittivity of free space, $Q_{l,s}'$ ($\mathrm{Cs^{-1}}$) is the charge production rate  on large/ small particles and $Q_{l,s}$ ($\mathrm{C}$) is the accumulated grain charge. In the tenuous, $\mathrm{CO_2}$-dominated Martian atmosphere, the dielectric constant is effectively unity; we therefore take $\epsilon \approx \epsilon_0 = 8.854 \times 10^{-12}\ \mathrm{F\ m^{-1}}$. 

The atmospheric conductivity $\sigma$ in a dust-free column is controlled by ion and electron production (galactic cosmic rays, solar UV) and by losses through ion–ion recombination and attachment to aerosols. Increased dust loading generally reduces the gas-phase conductivity by scavenging mobile ions; during global dust storms, reported values decrease from $\sim 10^{-12}$ to $\sim 8.1 \times 10^{-15}\ \mathrm{S\,m^{-1}}$ (\citealt{Sheel2025}).

However, detailed studies by \citealt{Cardnell2016JGRE..121.2335C}, using ionization and chemistry simulations regarding the lower Martian atmosphere, show that atmospheric conductivity can vary by several orders of magnitude across local time and dust conditions. 
Major drivers are daytime photoionization, nighttime ion-ion recombination, aerosol attachments as well as dust abundances throughout the atmospheric column. 
Their model predictions include an increasing $\sigma $ up to $10^{-9} \; \mathrm{Sm}^{-1}$ in daytime under strong photoionization, and decreases to $2 \cdot 10^{-13} \; \mathrm{Sm}^{-1}$ in nighttime or dust-rich, ion depleted conditions.
The charge relaxation timescale $\tau = \frac{\epsilon _0 }{ \sigma}$ is controlled by the conductivity, and sensitive to these variations. 
Because the steady-state grain charge in Eq.(~\ref{eq:tribo_charging_with_loss}) scales as 
$Q \propto \epsilon_0 / \sigma = \tau$,
higher conductivity as predicted by \citet{Cardnell2016JGRE..121.2335C} would substantially limit charge accumulation and reduce electric field magnitudes.
For example, $\sigma = 10^{-14} \; \mathrm{Sm}^{-1}$ yields $\tau \approx 900 \mathrm{s}$, while $\sigma = 10^{-10} \; \mathrm{Sm}^{-1}$ reduces $\tau$ to $\approx 0.009 \mathrm{s}$, effectively preventing any significant charge buildup.

This formulation highlights that atmospheric conductivity constitutes a first-order control on electric-field evolution through its regulation of the charge relaxation timescale $\tau = \epsilon_0 / \sigma$. Regions of high conductivity, such as the dayside upper atmosphere, are characterized by rapid charge dissipation and therefore inhibit the buildup of strong electric fields. Conversely, low-conductivity conditions, particularly in the nighttime lower atmosphere or in dust-rich ion-depleted regions, favor charge accumulation and enhance electric-field growth. Therefore, the predicted electric-field magnitudes must be interpreted in the context of the local conductivity regime, with the strongest fields occurring preferentially where turbulent charging coincides with reduced conductivity.

 \subsection{Electric field: Simulation Model}

To capture spatial charge separation, transport, and boundary effects, the system is described by Poisson’s equation coupled with the charge–continuity relation. The electrostatic potential $\phi$ satisfies

\begin{align}
\nabla \cdot E &= \frac{\rho}{\epsilon_0},
\qquad E = -\nabla \phi
\;\;\Rightarrow \;\; -\nabla^{2}\phi = \frac{\rho}{\epsilon_0},
\label{eq:poisson}
\end{align}

where $\rho$ ($\mathrm{Cm}^{-3}$) is the charge density and $E$ ($\mathrm{Vm}^{-1}$) is the electric field.
Charge conservation is expressed as

\begin{align}
\frac{\partial \rho}{\partial t} + \nabla \cdot J &= S,
\qquad J = \sigma E ,
\label{eq:electric field}
\end{align}

with ${J}$ ($\mathrm{Am}^{-2}$) the conduction current, $\sigma$ the atmospheric conductivity, and $S$ the collisional charging source. 

Using the triboelectric model of (\citet{Desch2000};\citet{Sheel2025}), $S$ can be written as the net contribution of collisions,

\begin{align}
S \approx n_l \nu_c \Delta q ,
\label{eq: charge source}
\end{align}

where $n_l$ ($\mathrm{m}^{-3}$)is the number density of large grains, $\nu_c$ their collisional frequency with small grains, and $\Delta q$ the per-collision charge exchange Eq.~(\ref{eq:charge_potential}). 
Thus, the triboelectric source term provides a simple link between microscopic collisional charging and macroscopic buildup of electric fields.

The triboelectric source term used here follows the standard approach employed in most previous Martian dust-devil and dust-storm charging studies (e.g., \citealt{Farrell1999}; \citealt{Desch2000}; \citealt{Sheel2025}). 

Recent in situ detections of electric discharges on Mars by \citet{Chide2025} indicate that discharge occurrence correlates with convective and turbulent structures, rather than with the bulk of high–opacity dust regions. 


At each model timestep $(\sim 15 min)$, the total charge distribution is computed using Eq.~(\ref{eq: charge source}). 
The electric potential is obtained by solving Eq.~(\ref{eq:poisson}) with a finite-difference scheme. Boundary conditions are imposed as follows: a Dirichlet condition at the lower boundary (Martian surface) with $\phi=0$, and a homogeneous Neumann (zero-gradient) condition at the upper boundary. The numerical implementation of the above equations is described in Appendix~\ref{app1}. \\

In contrast to earlier simulations of dust-devil electric fields, both small and large particle populations are explicitly charged and evolved self-consistently under the NASA Ames Mars GCM. No artificial separation of particle species is applied; their vertical distributions emerge from the model dynamics. The resulting number-density distributions of small and large particles at the onset of the GDS is shown in Fig.~\ref{fig:numberdensity distribution}.
Continuously resolving charge evolution between global timesteps would be computationally expensive, as it requires tracking the full motion of both particle populations across the planet at every substep. 
To retain accumulated charge at manageable cost, we advect a conserved charge-per-particle diagnostic. Specifically, at the end of each timestep we normalize the total charge in each grid cell by the local number densities of small and large particles; in the subsequent timestep, this charge per particle is rescaled using the updated particle distributions across the grid.
This approach preserves accumulated charging in a manner consistent with particle transport while keeping the computation manageable.


\section{Results}\label{sec:3}

Before discussing the electric-field magnitudes, we emphasize that in the present framework, charging efficiency is primarily governed by turbulence-driven collisional activity, while dust abundance acts as a necessary but not sufficient condition. As a result, enhanced electric fields arise where sufficient dust loading coincides with strong dynamical activity.

\subsection{Storm-time evolution of turbulence and conductivity}

\begin{figure}[!htbp]
    \centering
    \includegraphics[width=1\linewidth]{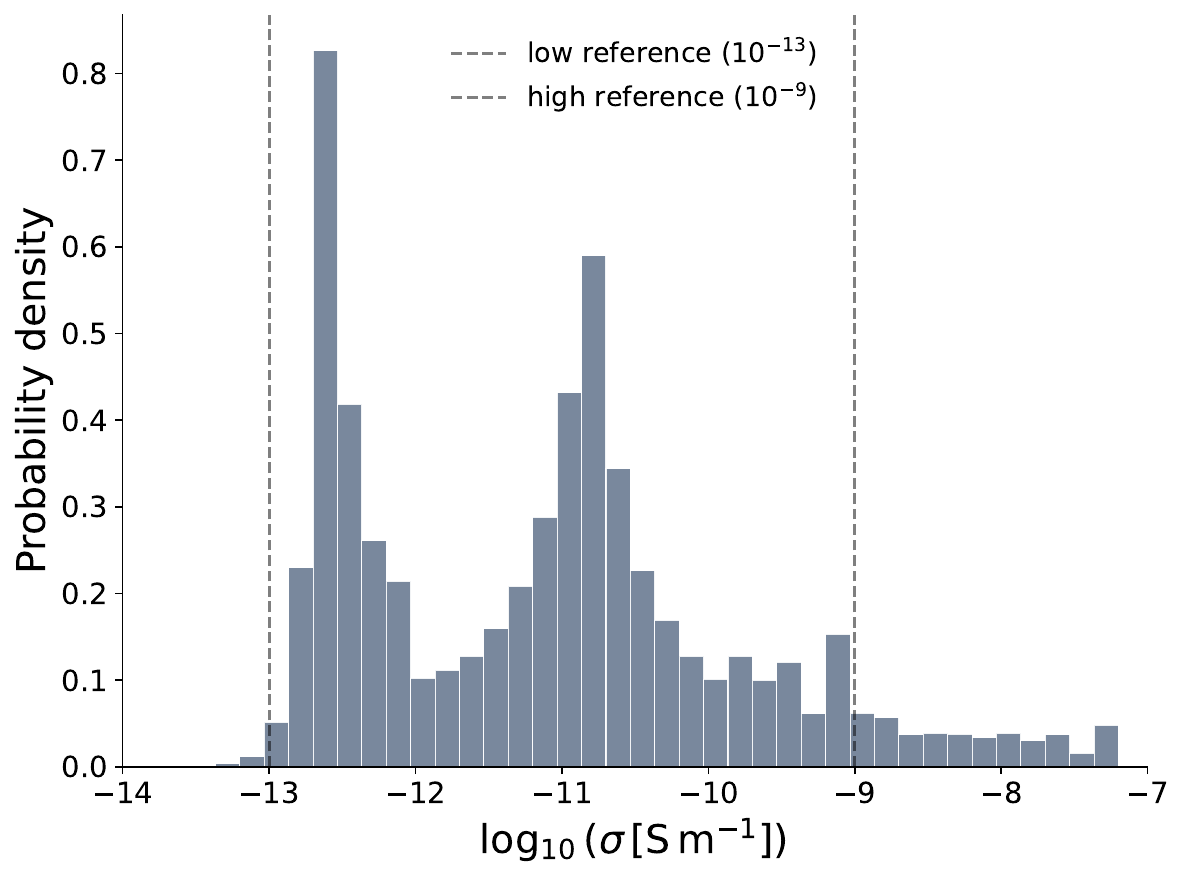}
    \caption{Distribution of parameterized atmospheric conductivity values, $\sigma $, sampled over the full simulation domain and time interval. The histogram summarizes the range of conductivities produced by the adopted parameterization. Vertical dashed lines indicate representative low- and high-conductivity reference values from \citet{Cardnell2016JGRE..121.2335C}, used to place the modeled charge-relaxation regime in context.}
    \label{fig: atmospheric conductivity histogramm}
\end{figure}

\begin{figure*}[!thtbp]
    \centering
    \includegraphics[width=1
    \linewidth]{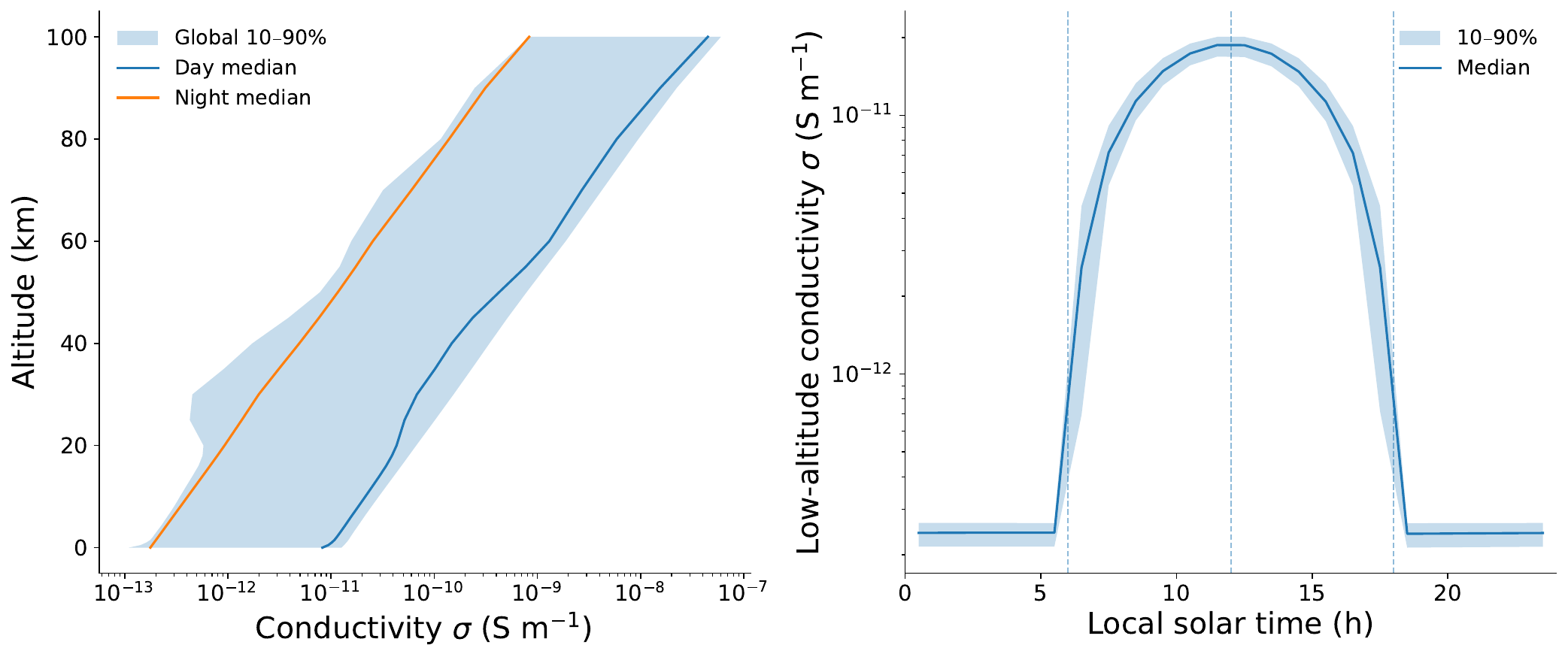}
    \caption{Parameterized atmospheric conductivity as a function of altitude and local solar time. Left: global 10--90\% range of $\sigma $ versus altitude, with separate day and night median profiles. Right: lower-atmosphere conductivity as a function of local solar time, shown as the median with 10--90\% variability. Together, these diagnostics illustrate the strong day-night modulation of charge-relaxation conditions assumed in the model.}
    \label{fig: atmosperic conductivity altitude}
\end{figure*}

The parameterized atmospheric conductivity varies by several orders of magnitude across the simulated storm, reflecting combined dependence on altitude, solar illumination, and dust loading. The background conductivity is represented by an exponential vertical profile with a dayside enhancement due to solar illumination, and is further modified by a dust-dependent factor intended to capture the competing effects of charge scavenging by aerosols and photoelectric electron production. 
Fig.~\ref{fig: atmospheric conductivity histogramm} summarizes the distribution of atmospheric conductivity values produced by the adopted parameterization across the simulation domain, together with representative low and high conductivity reference ranges by \citealt{Cardnell2016JGRE..121.2335C}. In Fig.~\ref{fig: atmosperic conductivity altitude} the  variation of the atmospheric conductivity in altitude and local solar time (day--night) differences are seen, providing the framework for the electric-field evolution discussed below.
These parameter choices are intentionally conservative, in the sense that they represent a moderate estimate of dust-induced conductivity variability and therefore avoid overstating charge-retention effects. 
The adopted scheme is intended primarily to capture the presently assumed first-order day-night contrast in Martian atmospheric conductivity, rather than to reproduce the full complexity of ionization chemistry.
Although this approach allows $\sigma $ to evolve in space and time, it remains a simplified but physically motivated representation of Martian atmospheric conductivity.

\subsection{Electric field }

\begin{figure}[!htbp]
    \centering
    \includegraphics[width=1\linewidth]{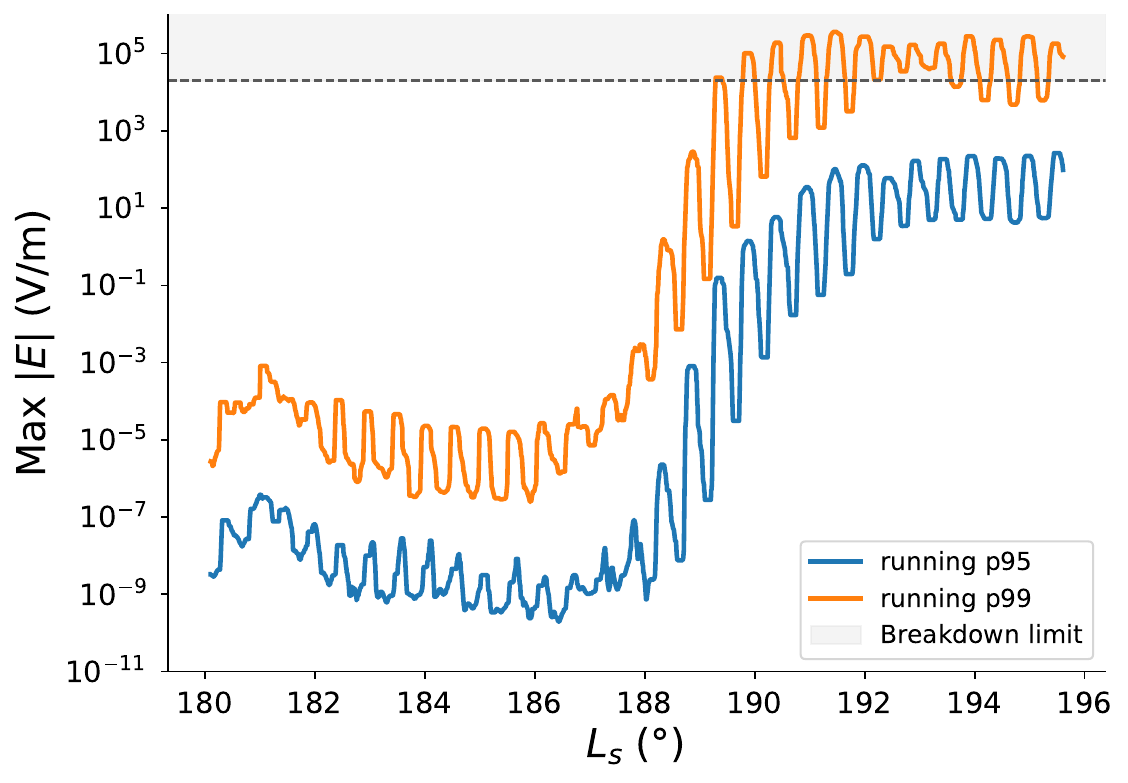}
    \caption{Running p95 (blue) and p99 (orange) electric-field magnitudes as a function of time. The discharge threshold (breakdown limit, $25\ \mathrm{kVm^{-1}}$) is shown in light gray. }
    \label{fig: maxE_all}
\end{figure}

As seen in Fig.~\ref{fig: maxE_all},  the upper tail of the electric-field distribution, represented by the running 95th and 99th percentiles (p95 and p99) of the electric-field magnitude distribution, remains very weak early in the simulation and then rises rapidly after $L_s \approx 187^{\circ}$, reaching several tens of $kVm^{-1}$. The adopted near-surface discharge threshold of $\sim 25\, kV/m$ is reached repeatedly after $L_s \approx 190 ^{\circ} $. These magnitudes are broadly comparable with previous dust-devil studies (e.g. \citet{Melnik1998,Farell2003,Delroy2011,Sheel2025}). 
Reporting electric fields of order $\approx 10^{4} - 10^{6}\; Vm^{-1}$, when charge-limiting processes are neglected, and field saturations around tens of $kVm^{-1}$ including charge saturation effects. 

\begin{figure}[!htbp]
    \centering
    \includegraphics[width=1\linewidth]{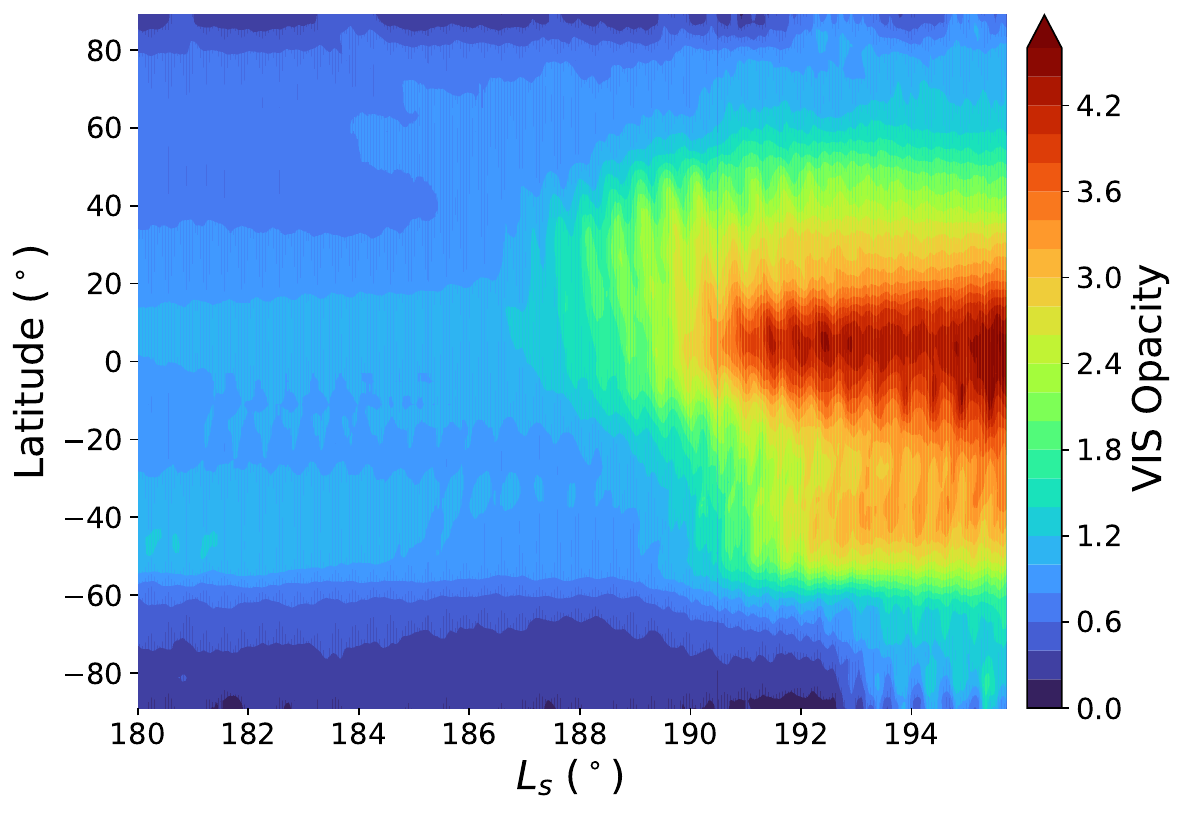}
    \caption{Visual opacity of the analyzed dataset, averaged over longitude, illustrating the build-up of the GDS. }
    \label{fig:VISOpacity}
\end{figure}

Up to $L_s \approx 187^\circ$, field magnitudes show no significant increase, and remain well below breakdown. 
Between $L_s \approx 187^\circ$ and $L_s \approx 191^\circ$,  fields strengthen during the phase of increasing dust loading consistent with the rising optical depth in Fig.~\ref{fig:VISOpacity}, which coincides with enhanced dynamical activity and increased collisional rates, consistent with the concurrent increase in collisional activity. 
Afterwards until $L_s \approx 195 ^{\circ} $ there is no significant change in the upper limit anymore. Because the dataset ends at this point, any inference about the later rise in opacity toward $\sim 5$ near $L_s=200^\circ$, and its possible effect on field strength, is speculative.

\begin{figure}[h!]
\centering
\includegraphics[scale=0.35, trim=0.8cm 0cm 0cm 0cm, clip]{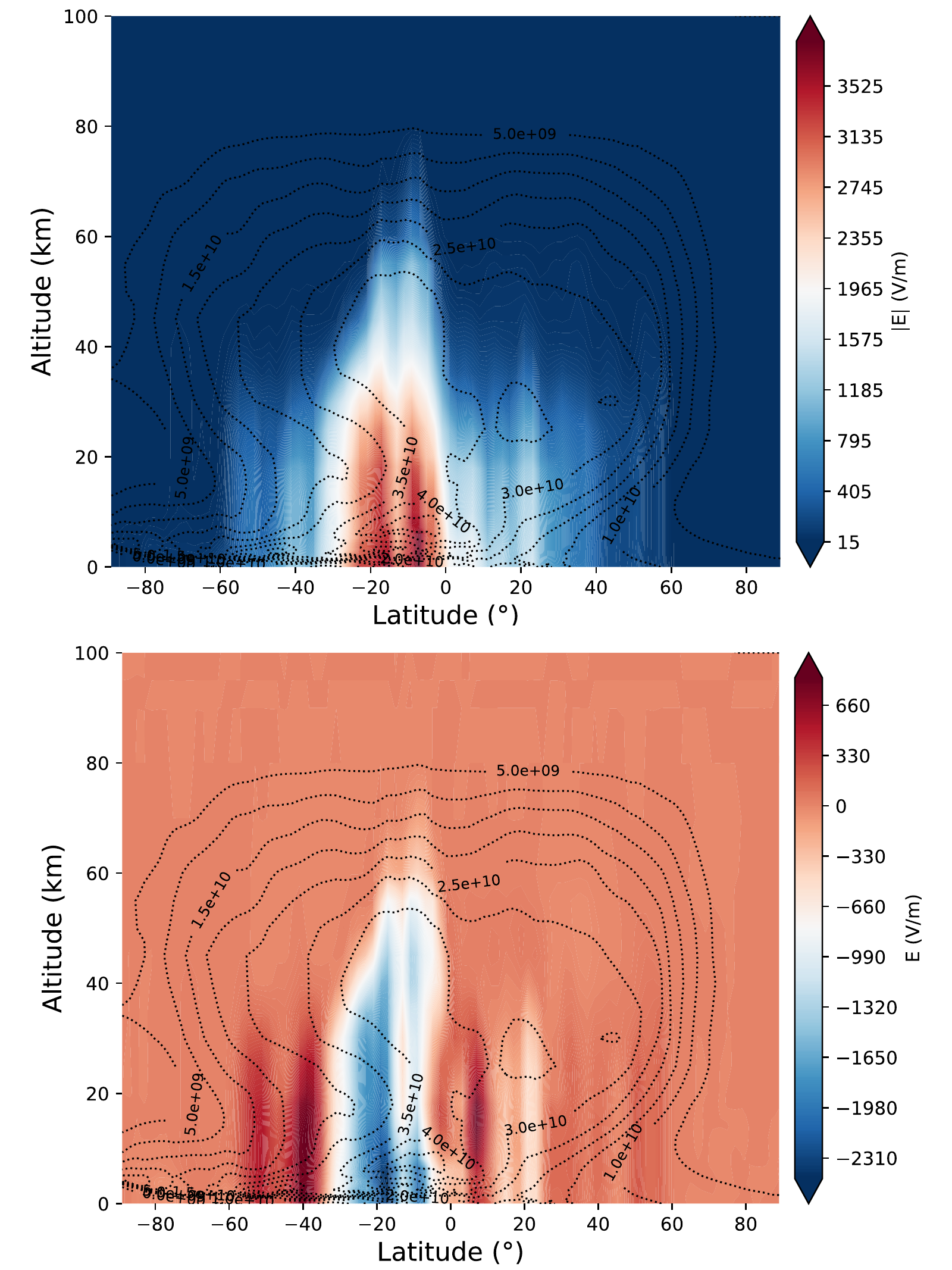}
\caption{Vertical slice through the planet at the final simulation timestep ($L_{\mathrm{s}} \sim 195^{\circ}$). Top: The longitudinally averaged electric field is shown. 
Bottom: Corresponding vertical slice showing the longitudinally averaged absolute electric-field magnitude.  In both panels are overlaid with contours indicating the total particle number density (small and large dust populations combined), from low (white) to high (red) values.
}
\label{fig:slice}
\end{figure}

The evolution shows regular oscillations, mirrored by oscillations in the planet-wide maxima of small and large particle number densities (Fig.~\ref{fig:numberdensity distribution}), indicating that part of the variability reflects changes in collisional charging frequency. At the same time, as discussed later in Sect.~\ref{sec: spatial extend}, the regular spacing of the peaks points to diurnal modulation: reduced nighttime conductivity leads to longer charge-relaxation times and thus favors electric-field buildup, whereas higher daytime conductivity promotes more rapid charge loss. Figure~\ref{fig:slice} illustrates the corresponding vertical structure of the electric field during the final phase of the analyzed GDS evolution. The strongest electric fields remain concentrated in the lower atmosphere, where dust loading and turbulence-driven collisional activity are most pronounced, while field magnitudes decrease rapidly with altitude.

\subsection{Spatial extent and diurnal organization}\label{sec: spatial extend}

\begin{figure}[!htbp]
    \centering
    \includegraphics[width=1\linewidth]{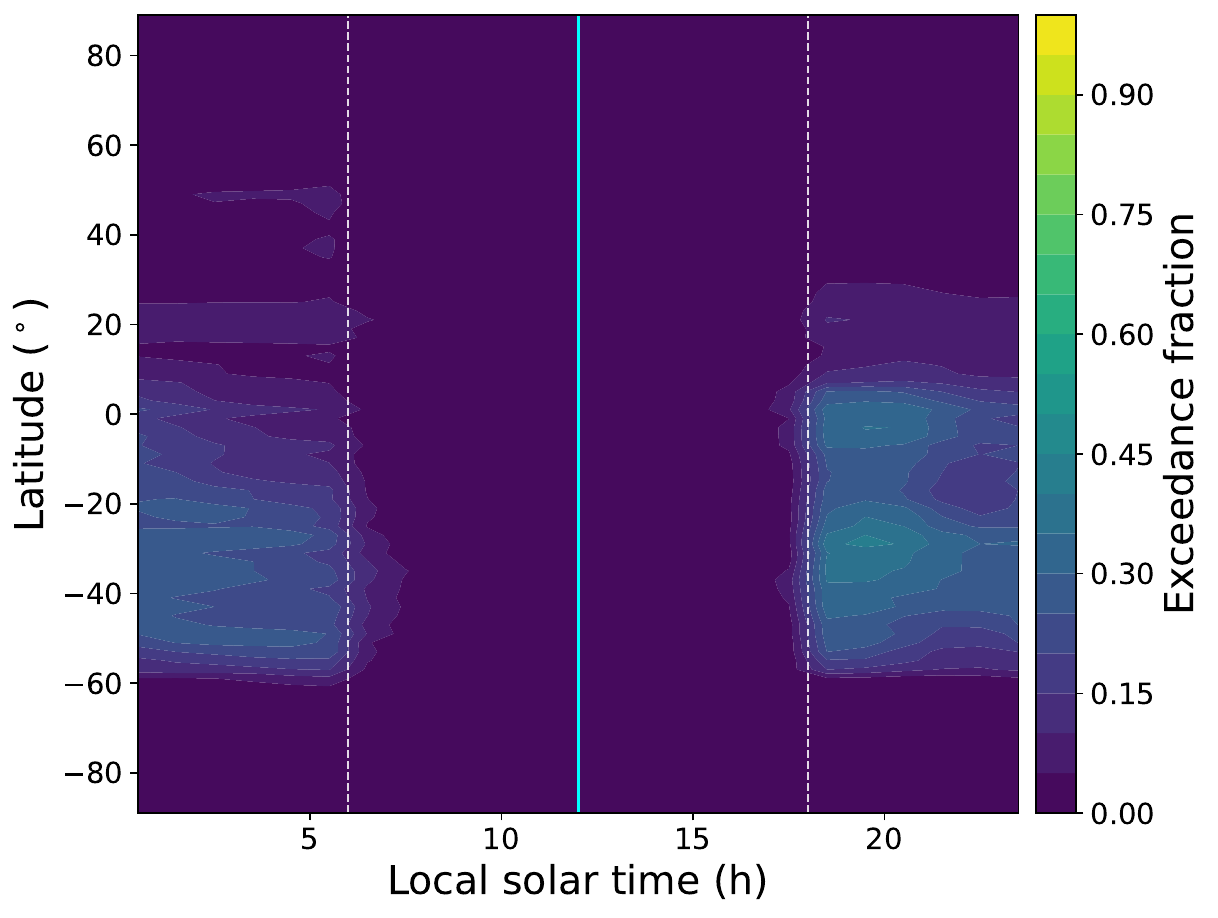}
    \caption{Local-time and latitude dependence of the exceedance fraction for discharge-relevant electric fields. Colors show the fraction of grid cells exceeding the adopted threshold as a function of local solar time and latitude, accumulated over the simulation period. Threshold exceedances are suppressed near local noon and are most frequent during evening and nighttime hours, especially in southern low-to-mid latitudes.}
    \label{fig: vertical slice2}
\end{figure}

To characterize where and when discharge-relevant electric activity occurs during the storm, we analyze the accumulated threshold exceedance as a function of geographic location and local solar time. Rather than reflecting only instantaneous field maxima, these diagnostics highlight the persistent spatiotemporal organization of electrical activity over the course of the GDS. The local-time dependence of the exceedance fraction (Fig.~\ref{fig: vertical slice2}) reveals a pronounced diurnal modulation. Exceedances are strongly suppressed near local noon, when enhanced daytime conductivity leads to rapid charge relaxation and inhibits large electric-field buildup, and increase again during the late afternoon, evening, and nighttime hours. The highest exceedance fractions are concentrated primarily in southern low-to-mid latitudes, extending from the equator to about $55^\circ$S. This pattern reflects the combined influence of turbulence-driven charge production and conductivity-dependent charge relaxation.

\begin{figure*}[!thtbp]
    \centering
    \includegraphics[width=0.9\linewidth]{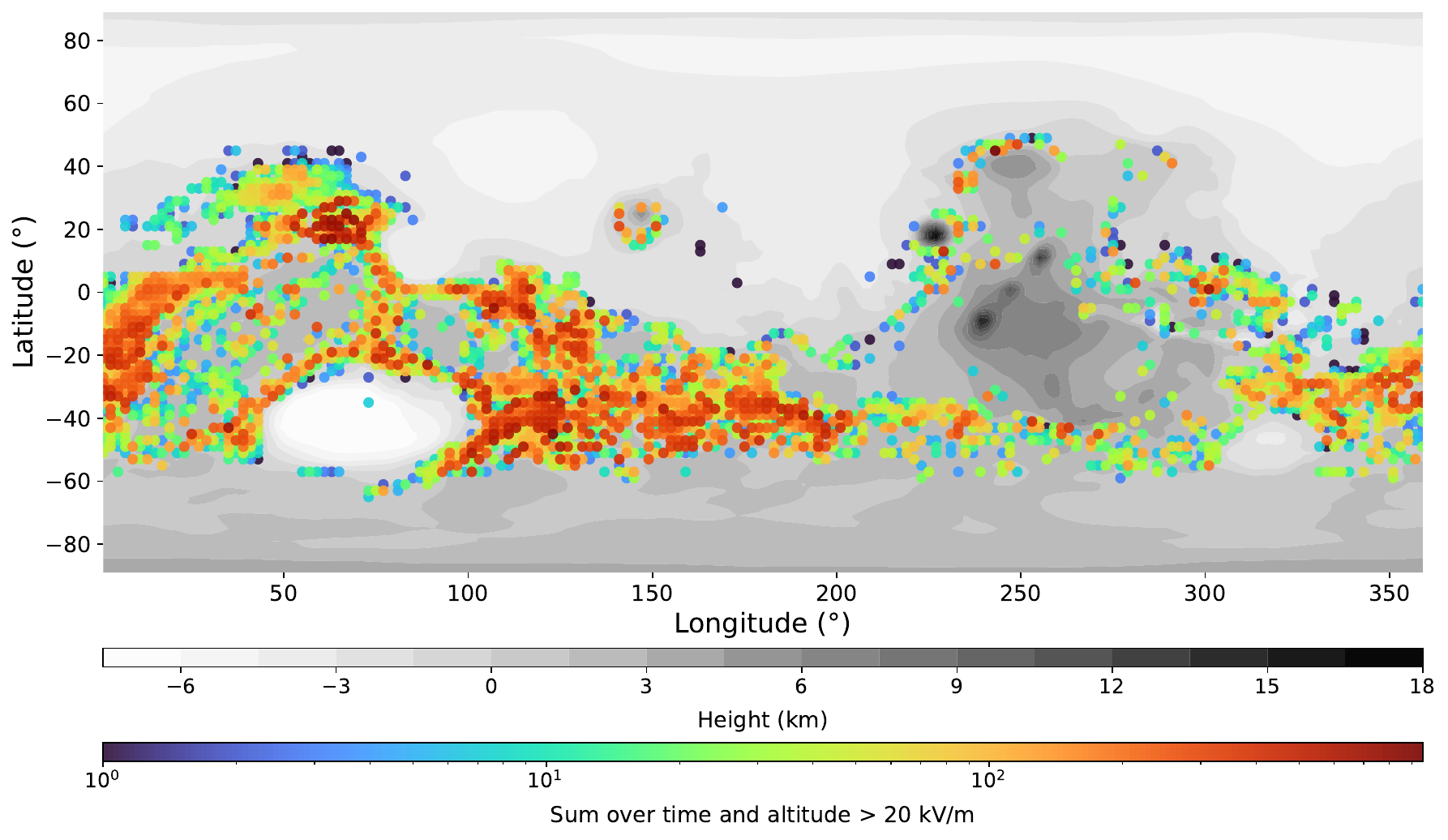}
    \caption{Spatial distribution of grid cells exceeding the discharge threshold, accumulated over the full simulation period and vertically integrated by altitude. Grey contours indicate the surface structures.}
    \label{fig:discharge_map}
\end{figure*}

Fig.~\ref{fig:discharge_map} shows the longitude–latitude distribution of discharge-threshold exceedances,
accumulated over time and vertically integrated over altitude. The most frequent
exceedances occur in the southern low-to-mid latitudes, approximately between  $0^\circ$ and $55^\circ$S, with additional regions of enhanced activity in northern
low-to-mid latitudes.

\begin{figure*}[!thtbp]
    \centering
    \includegraphics[width=0.9\linewidth]{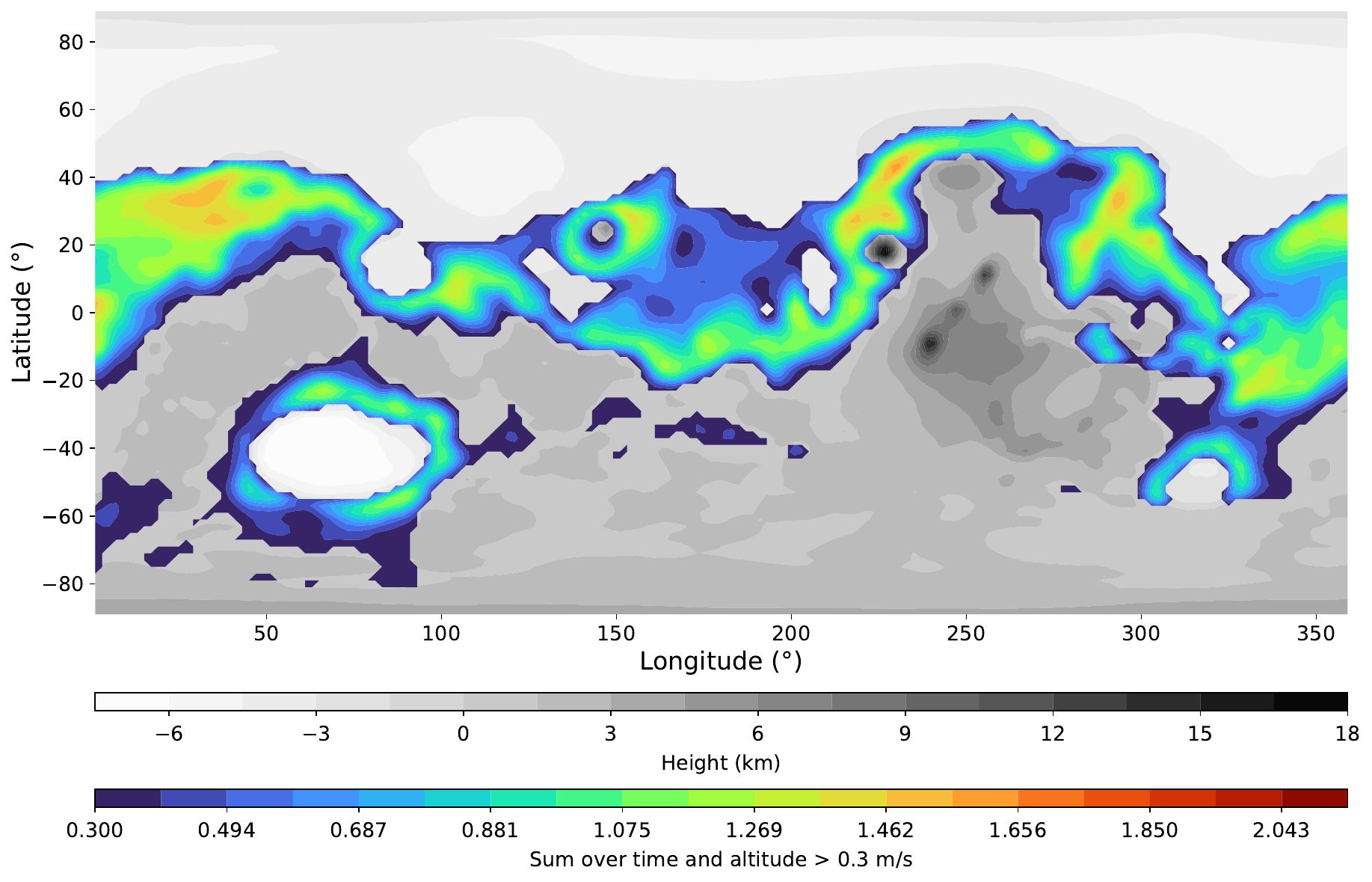}
    \caption{Spatial distribution of dynamically active regions, defined as grid cells exceeding a velocity threshold of 0.3 $\text{m s}^{-1}$, accumulated over the full simulation period and vertically integrated by altitude.}
    \label{fig:V_map}
\end{figure*}

To investigate the dynamical origin of this spatial organization, Fig.~\ref{fig:V_map} presents
the corresponding distribution of turbulent activity, expressed in terms of grid
cells exceeding a threshold velocity of $\sim$0.3$\text{m/s}$. We note that the exact threshold used to identify dynamically active regions does not affect the qualitative spatial correspondence, which remains robust across reasonable velocity ranges.  A clear spatial correspondence is observed
between regions of enhanced turbulence and the occurrence of discharge-relevant
electric fields. This supports the interpretation that electric-field generation
during global dust storms is primarily controlled by turbulence-driven collisional activity, rather than by dust abundance alone.

\subsection{Reduced field E/N}

\begin{figure}[!htbp]
    \centering
    \includegraphics[width=1\linewidth]{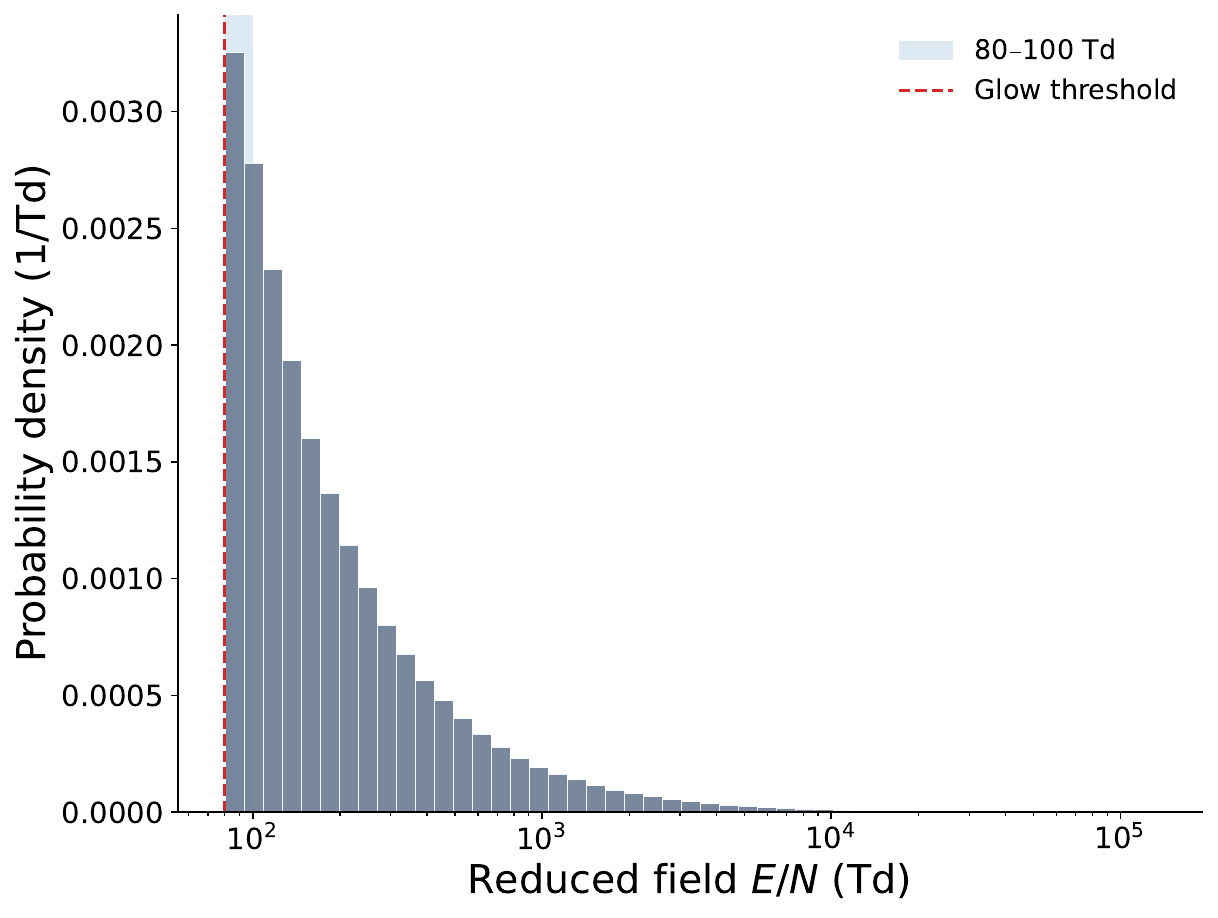}
    \caption{Relative-frequency histogram of the reduced field $E/N $ (Townsends) over all simulation times and global grid cells that met the glow-exceedance criterion in a $CO_2$ atmosphere. The y-axis shows bin probability (unitless). Dashed lines indicate the adopted discharge thresholds: red-glow onset (80 Td); green-streamer onset ($\sim$150 Td).}
    \label{fig:reduced field}
\end{figure}

We previously adopted a fixed discharge threshold appropriate for the near-surface atmosphere, as is common in dust-devil studies focused on the lowest layers immediately above the ground (\citet{Farrell1999, Barth2016Icar..268..253B}). In contrast, our simulations span several tens of kilometers in altitude, over which variations in neutral gas density become substantial. To account for this, we additionally evaluate the reduced electric field, $E/N$ (in Townsends, $1 \;Td = 10^{-21}\;  V m^{-2} $), which normalizes the electric field by the local neutral gas number density and is used for comparing discharge conditions across altitudes and environments (\citet{Qin2015GeoRL..42.2031Q, Radmilovi2021Atmos..12..438R}). 
Under Martian $CO_2$ conditions,  reduced fields of roughly $E/N \sim$ 70 to 100 Td are associated with the onset of ionization and sustainable glow-like discharge behavior (\citet{Guerra2022JAP...132g0902G, Liu2025PSST...34c5003L}).
Accordingly, we analyze the distribution of reduced fields for all simulation times and grid cells satisfying the glow criterion $E/N \geq 80\ \mathrm{Td} $ (Fig.~\ref{fig:reduced field}). 

The resulting histogram is strongly weighted toward values just above the glow threshold, with progressively fewer events at larger $E/N$, indicating that most discharge-relevant events remain weakly super-threshold. Therefore, extreme reduced fields, potentially compatible with streamer-like conditions, occur much less frequently.

\subsection{Altitude-dependent threshold exceedance and excess energy}

\begin{figure}[!htbp]
    \centering
    \includegraphics[width=1\linewidth]{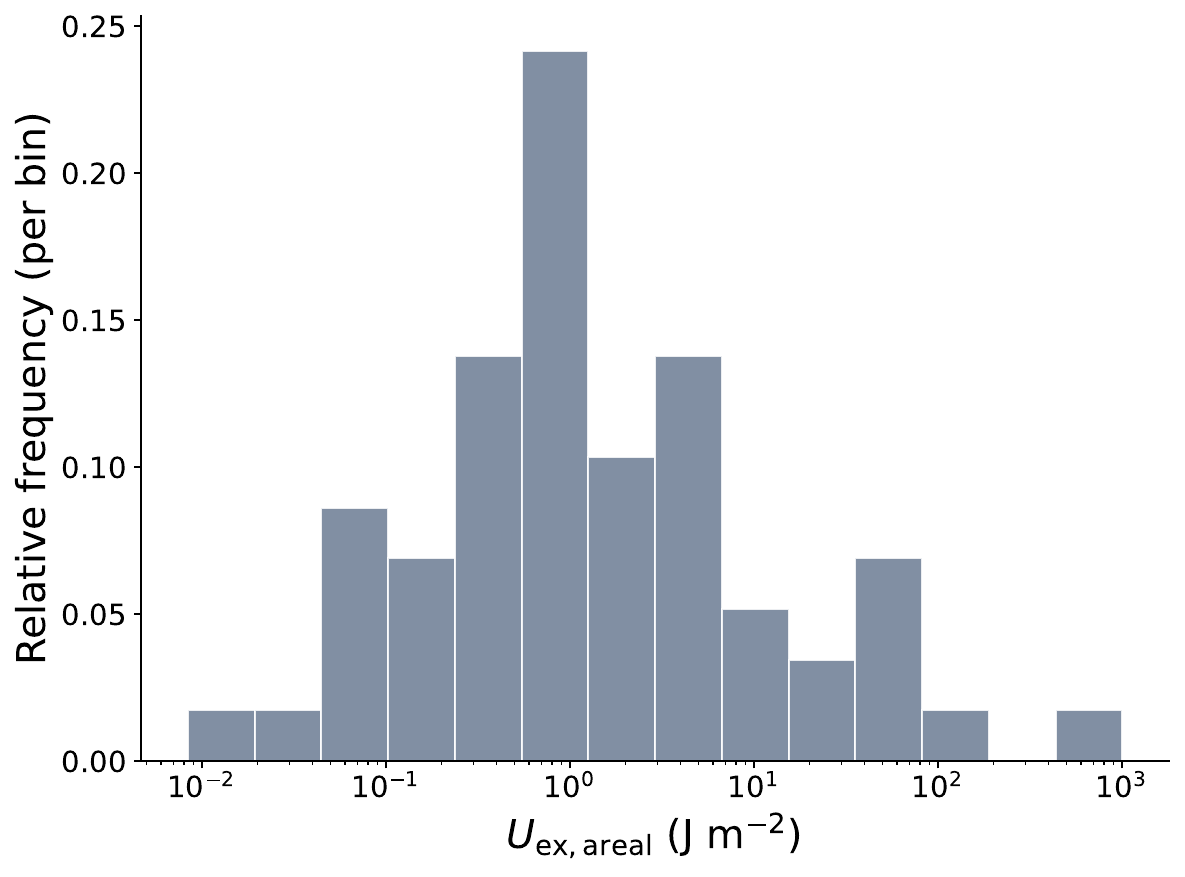}
    \caption{Frequency distribution of the column-integrated excess energy above the discharge threshold, $U_{ex, areal}$ ($Jm^{-2}$). Values are pooled over all output times and over all horizontal columns that exhibited $U_{ex, areal} \geq 0 $ at least once. }
    \label{fig:histogramm energy density}
\end{figure}

Fig.~\ref{fig:histogramm energy density} presents the distribution of column-integrated excess energy per unit area, aggregated across all model output times and grid columns. Following electrostatic energy–density formulations used by terrestrial thunderstorms (\citet{Marshall2002JGRD..107.4052M, Coleman2003JGRD..108.4298C, Bruning2013JAtS...70.4012B}), our estimator integrates the excess above the discharge limit. Only vertical columns that exceed the discharge threshold at least once at any altitude are included in the analysis.
The distribution peaks near $U_{ex,areal} \sim 10^{0}\; \text{Jm}^{-2}$, indicating that most exceedances lie only modestly above the local threshold (Fig.~\ref{fig:histogramm energy density}). This is consistent with weak glow- or Townsend-like discharge activity dominating the modeled energy release. 
The sparsely populated upper tail shows that stronger super-threshold events do occur, but they are comparatively rare. 
Earth-like leader lightning, which requires long-lived, hot, highly conducting channels, is unlikely in Mars’s thin $CO_2 $ atmosphere; the histogram’s shape (few very large exceedances and no broad plateau at extreme values) is consistent with this expectation.

\begin{figure}[!htbp]
    \centering
    \includegraphics[width=1\linewidth]{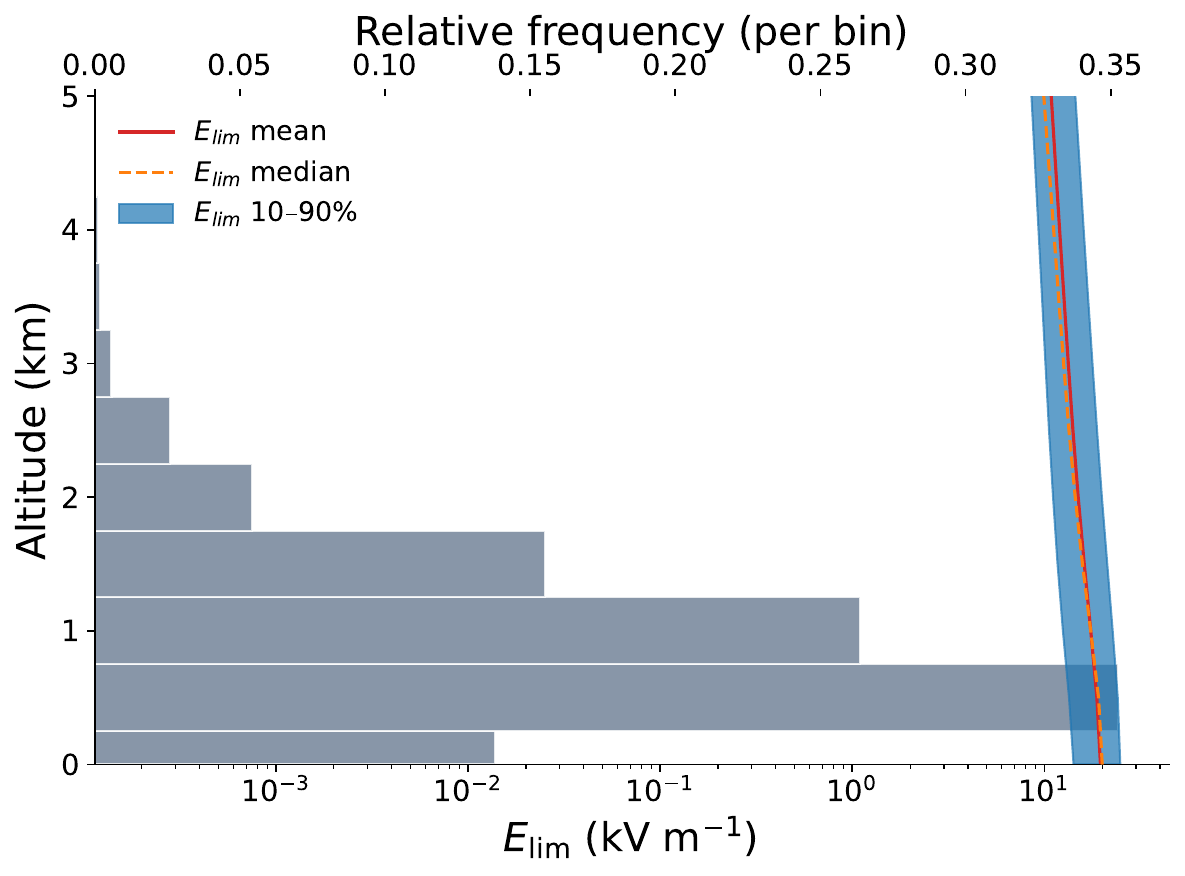}
    \caption{Horizontal bars show the relative frequency (bin probability; unitless) of occurrences with $E>E_{\mathrm{lim}}$ in each atmospheric layer, aggregated over the planet and all simulation steps. The upper x-axis reports bin probability. A thin overlay summarizes $E_{\mathrm{lim}}$ (mean, median, and 10–90\% range) as a function of altitude, with the bottom axis in kV\,m$^{-1}$.
    The overlaid curves indicate the altitude dependence of the discharge-limiting electric field $E_{\mathrm{lim}}$, derived from a constant reduced-field threshold $(E/N)_{\mathrm{th}}$ for CO$_2$ and anchored by a near-surface cap $E_{\mathrm{cap}}$ at reference density $N_{\mathrm{ref}}$.
    The shaded band shows the 10–90\,\% range across columns/times, with mean (solid) and median (dashed) curves.}
    \label{fig:histogramm energy density2}
\end{figure}

Fig.~\ref{fig:histogramm energy density2} shows the relative frequency of occurrences with $\mathrm{E > E_{lim}}$ as a function of altitude, integrated over all times and grid columns.
Although the limiting field $\mathrm{E_{lim}}$ decreases with altitude, exceedances remain most frequent near the surface, indicating that the vertical structure is controlled more by charging efficiency than by threshold reduction alone. 

\section{Discussion \& Conclusions} \label{sec:4}

This study investigates the occurrence and evolution of electric fields during planet-encircling dust storms in the Martian atmosphere. In contrast to previous work focused on dust devils, we examine the global circulation and analyze the buildup phase associated with dust lifting and its subsequent evolution. By coupling the model’s bimodal dust scheme with a triboelectric charging parameterization, we diagnose large-scale charging processes. Furthermore, we quantify the electric-field energy density and characterize discharge behavior, including onset thresholds, event frequency, and spatial clustering.

The simulated field intensities fall within expected ranges and are comparable to estimates for dust devils (\citealt{Farrell1999, Farrell2006, Delroy2011}). Field strength is primarily constrained by the relative velocities between small and large particles, atmospheric conductivity (which regulates the charge outflow timescale), particle number densities, and size contrasts. Globally, magnitudes span from hundreds of~$\mathrm{V\,m^{-1}}$ to several ~$10^{4}\ \mathrm{V\,m^{-1}}$ and exhibit pronounced spatial heterogeneity. The most frequent threshold exceedances occur within broad southern low-to-mid latitude belts, with secondary activity in northern low-to-mid latitudes; within these belts, some sectors show locally enhanced activity. The modeled electric field is predominantly downward directed, consistent with the smaller-scale dust-devil simulations of \citet{Sheel2022}, while localized opposite-signed plumes indicate sensitivity to regional meteorology, topography, and possibly soil composition.

A key result of this study is the strong spatial correspondence between regions of enhanced turbulent activity and the occurrence of discharge-relevant electric fields. As shown in Figs.~\ref{fig:discharge_map} and \ref{fig:V_map}, the geographic distribution of threshold exceedances closely follows the distribution of dynamically active regions characterized by elevated velocity fluctuations. This demonstrates that, at the global scale of dust storms, electric-field generation is primarily controlled by turbulence-driven collisional charging, rather than by dust abundance alone. In this framework, dust loading acts as a necessary but not sufficient condition, while local dynamical forcing determines where electrical activity is actually triggered.

Overall, the excess energy density above the discharge threshold is small, indicating that discharges are primarily in the Townsend or glow regime rather than Earth-like leader lightning. The upper tail of the distribution nonetheless allows for transient streamer activity during the most intense episodes. Given the pronounced spatial hotspots in the global distribution of electrical activity, targeted observations of these regions during future GDS events would be especially valuable.

Recent in-situ observations by \citet{Chide2025} indicate that electrical discharges on Mars occur predominantly within locally and dynamically active structures, such as dust devils, storm fronts, and sharp shear boundaries. This contrasts with earlier modeling approaches in which electric-field generation was primarily scaled with dust number density under the assumption of constant interparticle velocities. The results presented here provide a global-scale theoretical framework consistent with these observations, demonstrating that turbulent and convective dynamics are the primary drivers of electrification. Although global dust storms exhibit high dust abundances and large optical depths, their interiors become increasingly radiatively stabilized over time, suppressing vertical mixing and reducing turbulent motions (\citealt{Haberle2019}; \citealt{Montabone2020JGRE..12506111M}).

While the present model incorporates a turbulence-dependent charging parameterization, the resolved dynamical fields of the GCM remain inherently smooth and do not explicitly capture small-scale intermittent structures such as dust-devil vortices or sharp storm fronts. As a result, the simulated electric-field activity is distributed over broader regions than suggested by in situ observations. The model therefore captures the large-scale envelope of electrically favorable conditions, rather than the true spatial localization of discharge events. In reality, electrical activity is expected to be concentrated within subgrid-scale turbulent structures embedded within these broader regions. Consequently, the predicted spatial extent and frequency of discharge-prone areas should be interpreted as an upper bound.

Comparative analyses of electric-field buildup across multiple global dust storms would help refine regional attribution and test the robustness of spatial hotspots. A multi-storm, multi-year analysis could determine whether electrical activity in the lower Martian atmosphere consistently peaks in the same regions or varies interannually. Although the current framework captures the essential coupling between dust dynamics and triboelectric charging, further refinement—particularly through explicit plasma coupling and improved microphysical charge-transfer modeling—will enhance the fidelity of predicted field magnitudes. A key priority for future work is the incorporation of more realistic and explicitly resolved interparticle-velocity distributions and conductivity fields beyond the current proxy-based parameterizations. Although these additions come with increased computational expense, they would allow the model to better capture the spatial intermittency and temporal variability indicated by recent in-situ observations, and would substantially reduce uncertainties in predicted electric-field strengths. Such developments will enable more accurate assessments of Martian atmospheric electrification and its implications for surface operations, atmospheric chemistry, and future mission design.

\begin{acknowledgements}

The research presented here originated as the master’s thesis project of Ina Taxis within the Master’s Programme in Astrophysics at Ludwig-Maximilians-Universität (LMU) Munich.  L.G acknowledges financial support from the Severo Ochoa grant CEX2021-001131-S funded by MCIN/AEI/10.13039/501100011033 and 
Ministerio de Ciencia e Innovación through the project PID2022-137241NB-C43.  R.A.U., M.A.K., and A.S.B. acknowledge support from NASA’s Solar System Workings Program (grant number 80NSSC21M0124) through the Bay Area Environmental Research Institute, NASA’s Science Mission Directorate, Planetary Science Division, Mars Exploration Program, Mars Climate Modeling Center, and the NASA Advanced Supercomputing Division.
\end{acknowledgements}

\newpage

\bibliographystyle{aa}

\bibliography{ref}

\begin{thebibliography}{45}
\expandafter\ifx\csname natexlab\endcsname\relax\def\natexlab#1{#1}\fi

\bibitem[{{Abdelaal} {et~al.}(2025){Abdelaal}, {Dokuchaev}, {Kuznetsov},
  {Shashkova}, {Lyash}, {Dubov}, {Obod}, {Kartasheva}, {Dolnikov}, \&
  {Zakharov}}]{Abdelaal2025SoSyR..59...71A}
{Abdelaal}, M.~E., {Dokuchaev}, I.~V., {Kuznetsov}, I.~A., {et~al.} 2025, Solar
  System Research, 59, 71

\bibitem[{{Adzhieva} {et~al.}(2020){Adzhieva}, {Shapovalov}, {Mashukov}, \&
  {Tumgoeva}}]{Adzhieva2020JPhCS1604a2016A}
{Adzhieva}, A.~A., {Shapovalov}, V.~A., {Mashukov}, I.~K., \& {Tumgoeva}, H.~A.
  2020, in Journal of Physics Conference Series, Vol. 1604, Journal of Physics
  Conference Series, 012016

\bibitem[{{Aoki} {et~al.}(2022){Aoki}, {Gkouvelis}, {G{\'e}rard}, {Soret},
  {Hubert}, {Lopez-Valverde}, {Gonz{\'a}lez-Galindo}, {Sagawa}, {Thomas},
  {Ristic}, {Willame}, {Depiesse}, {Mason}, {Patel}, {Bellucci},
  {Lopez-Moreno}, {Daerden}, \& {Vandaele}}]{Aoki2022}
{Aoki}, S., {Gkouvelis}, L., {G{\'e}rard}, J.-C., {et~al.} 2022, Journal of
  Geophysical Research (Planets), 127, e07206

\bibitem[{{Barth} {et~al.}(2016){Barth}, {Farrell}, \&
  {Rafkin}}]{Barth2016Icar..268..253B}
{Barth}, E.~L., {Farrell}, W.~M., \& {Rafkin}, S. C.~R. 2016, \icarus, 268, 253

\bibitem[{{Bertrand} {et~al.}(2020){Bertrand}, {Wilson}, {Kahre}, {Urata}, \&
  {Kling}}]{Bertrand2020}
{Bertrand}, T., {Wilson}, R.~J., {Kahre}, M.~A., {Urata}, R., \& {Kling}, A.
  2020, Journal of Geophysical Research (Planets), 125, e06122

\bibitem[{{Bruning} \& {MacGorman}(2013)}]{Bruning2013JAtS...70.4012B}
{Bruning}, E.~C. \& {MacGorman}, D.~R. 2013, Journal of the Atmospheric
  Sciences, 70, 4012

\bibitem[{{Cardnell} {et~al.}(2016){Cardnell}, {Witasse}, {Molina-Cuberos},
  {Michael}, {Tripathi}, {D{\'e}prez}, {Montmessin}, \&
  {O'Brien}}]{Cardnell2016JGRE..121.2335C}
{Cardnell}, S., {Witasse}, O., {Molina-Cuberos}, G.~J., {et~al.} 2016, Journal
  of Geophysical Research (Planets), 121, 2335

\bibitem[{Chide {et~al.}(2025)Chide, Lorenz, Montmessin, Maurice, Parot, Hueso,
  Martinez, Vicente-Retortillo, Jacob, Lemmon, Dubois, Meslin, Newman,
  Bertrand, Deprez, Toledo, S{\'a}nchez-Lavega, Cousin, \& Wiens}]{Chide2025}
Chide, B., Lorenz, R.~D., Montmessin, F., {et~al.} 2025, Nature, 647, 865

\bibitem[{{Coleman} {et~al.}(2003){Coleman}, {Marshall}, {Stolzenburg},
  {Hamlin}, {Krehbiel}, {Rison}, \& {Thomas}}]{Coleman2003JGRD..108.4298C}
{Coleman}, L.~M., {Marshall}, T.~C., {Stolzenburg}, M., {et~al.} 2003, Journal
  of Geophysical Research (Atmospheres), 108, 4298

\bibitem[{{Delory} \& {Farrell}(2011)}]{Delroy2011}
{Delory}, G. \& {Farrell}, W. 2011, in EPSC-DPS Joint Meeting 2011, Vol. 2011,
  1229

\bibitem[{{Desch} \& {Cuzzi}(2000)}]{Desch2000}
{Desch}, S.~J. \& {Cuzzi}, J.~N. 2000, \icarus, 143, 87

\bibitem[{{Farrell} {et~al.}(2003){Farrell}, {Delory}, {Cummer}, \&
  {Marshall}}]{Farell2003}
{Farrell}, W.~M., {Delory}, G.~T., {Cummer}, S.~A., \& {Marshall}, J.~R. 2003,
  \grl, 30, 2050

\bibitem[{{Farrell} {et~al.}(1999){Farrell}, {Kaiser}, {Desch}, {Houser},
  {Cummer}, {Wilt}, \& {Landis}}]{Farrell1999}
{Farrell}, W.~M., {Kaiser}, M.~L., {Desch}, M.~D., {et~al.} 1999, \jgr, 104,
  3795

\bibitem[{{Farrell} {et~al.}(2006){Farrell}, {Renno}, {Delory}, {Cummer}, \&
  {Marshall}}]{Farrell2006}
{Farrell}, W.~M., {Renno}, N., {Delory}, G.~T., {Cummer}, S.~A., \& {Marshall},
  J.~R. 2006, Journal of Geophysical Research (Planets), 111, E01006

\bibitem[{{Forward} {et~al.}(2009){Forward}, {Lacks}, \&
  {Sankaran}}]{Forward2009GeoRL..3613201F}
{Forward}, K.~M., {Lacks}, D.~J., \& {Sankaran}, R.~M. 2009, \grl, 36, L13201

\bibitem[{{Franzese} {et~al.}(2018){Franzese}, {Esposito}, {Lorenz},
  {Silvestro}, {Popa}, {Molinaro}, {Cozzolino}, {Molfese}, {Marty}, \&
  {Deniskina}}]{Franzese2018}
{Franzese}, G., {Esposito}, F., {Lorenz}, R., {et~al.} 2018, Earth and
  Planetary Science Letters, 493, 71

\bibitem[{{Gkouvelis} {et~al.}(2020{\natexlab{a}}){Gkouvelis}, {G{\'e}rard},
  {Gonz{\'a}lez-Galindo}, {Hubert}, \& {Schneider}}]{Gkouvelis2020}
{Gkouvelis}, L., {G{\'e}rard}, J.~C., {Gonz{\'a}lez-Galindo}, F., {Hubert}, B.,
  \& {Schneider}, N.~M. 2020{\natexlab{a}}, \grl, 47, e87468

\bibitem[{{Gkouvelis} {et~al.}(2020{\natexlab{b}}){Gkouvelis}, {G{\'e}rard},
  {Ritter}, {Hubert}, {Schneider}, \& {Jain}}]{Gkouvelis2020b}
{Gkouvelis}, L., {G{\'e}rard}, J.-C., {Ritter}, B., {et~al.}
  2020{\natexlab{b}}, \icarus, 341, 113666

\bibitem[{{Guerra} {et~al.}(2022){Guerra}, {Silva}, {Pinh{\~a}o}, {Guaitella},
  {Guerra-Garcia}, {Peeters}, {Tsampas}, \& {van de
  Sanden}}]{Guerra2022JAP...132g0902G}
{Guerra}, V., {Silva}, T., {Pinh{\~a}o}, N., {et~al.} 2022, Journal of Applied
  Physics, 132, 070902

\bibitem[{{Haberle} {et~al.}(2019){Haberle}, {Kahre}, {Hollingsworth},
  {Montmessin}, {Wilson}, {Urata}, {Brecht}, {Wolff}, {Kling}, \&
  {Schaeffer}}]{Haberle2019}
{Haberle}, R.~M., {Kahre}, M.~A., {Hollingsworth}, J.~L., {et~al.} 2019,
  \icarus, 333, 130

\bibitem[{Harris {et~al.}(2021)Harris, Chen, Putman, Zhou, \&
  Chen}]{harris2021fv3}
Harris, L., Chen, X., Putman, W., Zhou, L., \& Chen, J.-H. 2021, A Scientific
  Description of the GFDL Finite-Volume Cubed-Sphere Dynamical Core (FV3),
  Tech. rep., NOAA Geophysical Fluid Dynamics Laboratory, Princeton, NJ, gFDL
  Technical Report

\bibitem[{{Harrison} {et~al.}(2016){Harrison}, {Barth}, {Esposito}, {Merrison},
  {Montmessin}, {Aplin}, {Borlina}, {Berthelier}, {D{\'e}prez}, {Farrell},
  {Houghton}, {Renno}, {Nicoll}, {Tripathi}, \&
  {Zimmerman}}]{Harrison2016SSRv..203..299H}
{Harrison}, R.~G., {Barth}, E., {Esposito}, F., {et~al.} 2016, \ssr, 203, 299

\bibitem[{Harrison {et~al.}(2017)Harrison, Barth, Esposito, Merrison,
  Montmessin, Aplin, Borlina, Berthelier, D{\'e}prez, Farrell, Houghton,
  Renn{\'o}, Nicoll, Tripathi, \& Zimmerman}]{HarrisonEtAl2017}
Harrison, R.~G., Barth, E., Esposito, F., {et~al.} 2017, in Space Sciences
  Series of ISSI, Vol.~59, Dust Devils, ed. D.~Reiss, R.~Lorenz, M.~Balme,
  L.~Neakrase, A.~P. Rossi, A.~Spiga, \& J.~Zarnecki (Springer), 299--345

\bibitem[{{Izvekova} \& {Popel}(2020)}]{Izvekova2020JPhCS1556a2071I}
{Izvekova}, Y.~N. \& {Popel}, S.~I. 2020, in Journal of Physics Conference
  Series, Vol. 1556, Journal of Physics Conference Series, 012071

\bibitem[{{Kahre} {et~al.}(2023){Kahre}, {Haberle}, {Wilson}, {Urata},
  {Steakley}, {Brecht}, {Bertrand}, {Kling}, {Batterson}, {Hartwick}, {Harman},
  \& {Gkouvelis}}]{Kahre2023}
{Kahre}, M.~A., {Haberle}, R.~M., {Wilson}, R.~J., {et~al.} 2023, \icarus, 400,
  115561

\bibitem[{{Kim} {et~al.}(2013){Kim}, {Wo}, {Maity}, {Atreya}, \&
  {Kaiser}}]{Kim2013JAChS.135.4910K}
{Kim}, Y.~S., {Wo}, K.~P., {Maity}, S., {Atreya}, S.~K., \& {Kaiser}, R.~I.
  2013, Journal of the American Chemical Society, 135, 4910

\bibitem[{{Kok} \& {Renno}(2006)}]{Kok2006GeoRL..3319S10K}
{Kok}, J.~F. \& {Renno}, N.~O. 2006, \grl, 33, L19S10

\bibitem[{{Liu} {et~al.}(2025){Liu}, {Silva}, {Dias}, {Viegas}, {Zhao}, {Du},
  {He}, \& {Guerra}}]{Liu2025PSST...34c5003L}
{Liu}, Y., {Silva}, T., {Dias}, T.~C., {et~al.} 2025, Plasma Sources Science
  Technology, 34, 035003

\bibitem[{{Marshall} \& {Stolzenburg}(2002)}]{Marshall2002JGRD..107.4052M}
{Marshall}, T.~C. \& {Stolzenburg}, M. 2002, Journal of Geophysical Research
  (Atmospheres), 107, 4052

\bibitem[{{Melnik} \& {Parrot}(1998)}]{Melnik1998}
{Melnik}, O. \& {Parrot}, M. 1998, \jgr, 103, 29107

\bibitem[{{Montabone} {et~al.}(2020){Montabone}, {Spiga}, {Kass},
  {Kleinb{\"o}hl}, {Forget}, \& {Millour}}]{Montabone2020JGRE..12506111M}
{Montabone}, L., {Spiga}, A., {Kass}, D.~M., {et~al.} 2020, Journal of
  Geophysical Research (Planets), 125, e06111

\bibitem[{{Navarro-Gonz{\'a}lez} {et~al.}(2010){Navarro-Gonz{\'a}lez},
  {Vargas}, {de la Rosa}, {Raga}, \& {McKay}}]{Navarro2010JGRE..11512010N}
{Navarro-Gonz{\'a}lez}, R., {Vargas}, E., {de la Rosa}, J., {Raga}, A.~C., \&
  {McKay}, C.~P. 2010, Journal of Geophysical Research (Planets), 115, E12010

\bibitem[{{Putman} \& {Lin}(2007)}]{Putman2007JCoPh.227...55P}
{Putman}, W.~M. \& {Lin}, S.-J. 2007, Journal of Computational Physics, 227, 55

\bibitem[{{Qin} \& {Pasko}(2015)}]{Qin2015GeoRL..42.2031Q}
{Qin}, J. \& {Pasko}, V.~P. 2015, \grl, 42, 2031

\bibitem[{{Radmilovi{\'c}-Radjenovi{\'c}}
  {et~al.}(2021){Radmilovi{\'c}-Radjenovi{\'c}}, {Sabo}, \&
  {Radjenovi{\'c}}}]{Radmilovi2021Atmos..12..438R}
{Radmilovi{\'c}-Radjenovi{\'c}}, M., {Sabo}, M., \& {Radjenovi{\'c}}, B. 2021,
  Atmosphere, 12, 438

\bibitem[{{Ruf} {et~al.}(2009){Ruf}, {Renno}, {Kok}, {Bandelier}, {Sander},
  {Gross}, {Skjerve}, \& {Cantor}}]{Ruf2009GeoRL..3613202R}
{Ruf}, C., {Renno}, N.~O., {Kok}, J.~F., {et~al.} 2009, \grl, 36, L13202

\bibitem[{Sheel \& Haider(2016)}]{Sheel2016}
Sheel, V. \& Haider, S.~A. 2016, Journal of Geophysical Research: Space
  Physics, 121, 8038

\bibitem[{{Sheel} \& {Uttam}(2022)}]{Sheel2022}
{Sheel}, V. \& {Uttam}, S. 2022, in 44th COSPAR Scientific Assembly. Held 16-24
  July, Vol.~44, 381

\bibitem[{{Sheel} {et~al.}(2025){Sheel}, {Uttam}, \& {Mishra}}]{Sheel2025}
{Sheel}, V., {Uttam}, S., \& {Mishra}, S.~K. 2025, Physics of Plasmas, 32,
  033704

\bibitem[{Smith \& Guzewich(2019)}]{Smith2019}
Smith, M.~D. \& Guzewich, S.~D. 2019, in 49th International Conference on
  Environmental Systems (ICES), International Conference on Environmental
  Systems, Boston, Massachusetts, USA, paper ICES-2019-181

\bibitem[{{Soret} {et~al.}(2022){Soret}, {G{\'e}rard}, {Aoki}, {Gkouvelis},
  {Thomas}, {Ristic}, {Hubert}, {Willame}, {Depiesse}, {Vandaele}, {Patel},
  {Mason}, {Daerden}, {L{\'o}pez-Moreno}, \& {Bellucci}}]{Soret2022}
{Soret}, L., {G{\'e}rard}, J.-C., {Aoki}, S., {et~al.} 2022, Journal of
  Geophysical Research (Planets), 127, e07220

\bibitem[{{Soret} {et~al.}(2025){Soret}, {Robin}, {G{\'e}rard}, {Gkouvelis},
  {Thomas}, {Ristic}, {Willame}, {Hubert}, {Vandaele}, {Mason}, {Daerden}, \&
  {Patel}}]{Soret2025}
{Soret}, L., {Robin}, H., {G{\'e}rard}, J.-C., {et~al.} 2025, \icarus, 441,
  116707

\bibitem[{{Urata} {et~al.}(2024){Urata}, {Bertrand}, {Kahre}, {Wilson},
  {Kling}, \& {Wolff}}]{Urata2024LPICo3007.3337U}
{Urata}, R.~A., {Bertrand}, T., {Kahre}, M.~A., {et~al.} 2024, in LPI
  Contributions, Vol. 3007, Tenth International Conference on Mars, 3337

\bibitem[{{Urata} {et~al.}(2025){Urata}, {Bertrand}, {Kahre}, {Wilson},
  {Kling}, \& {Wolff}}]{Urata2025Icar..42916446U}
{Urata}, R.~A., {Bertrand}, T., {Kahre}, M.~A., {et~al.} 2025, \icarus, 429,
  116446

\bibitem[{{Zhang} \& {Zhou}(2020)}]{Zhang2020NatCo..11.5072Z}
{Zhang}, H. \& {Zhou}, Y.-H. 2020, Nature Communications, 11, 5072

\end{thebibliography}

\begin{appendix}

\section{Electric-Field}
\label{app1}

\subsection{Quasi-static 1-D Approach}

Charge separation and electric field buildup are treated column by column, solving Poisson’s equation in the vertical direction:

\begin{equation}
E(z,t) = - \frac{\partial \phi}{\partial z}, \qquad
\frac{\partial^2 \phi(z,t)}{\partial z^2} = -\frac{\rho(z,t)}{\epsilon_0}.
\end{equation}

Because charge transport in dust storms occurs at wind speeds far below the speed of light, the displacement current can be neglected, and the electric field is assumed to respond instantaneously to charge redistribution. This reduces memory and computational requirements while retaining physical fidelity.

\subsection{Finite difference discretization}

The Poisson equation is discretized using finite-difference stencils. On a uniform grid with spacing $\Delta z$, the discrete form is

\begin{equation}
\phi_{i-1}^{(n)} - 2\phi_i^{(n)} + \phi_{i+1}^{(n)}
= - \frac{\rho_i^{(n)} \Delta z^2}{\epsilon_0},
\end{equation}

with the electric field obtained as

\begin{equation}
E_i^{(n)} \approx - \frac{\phi_{i+1}^{(n)} - \phi_{i-1}^{(n)}}{2\Delta z}.
\end{equation}

Since the NASA Ames GCM employs altitude levels with non-uniform spacing, these expressions are generalized. Defining

\begin{align}
\Delta z_{\text{up},i} = z_{i+1}-z_i, \; \;
\Delta z_{\text{dn},i}=z_i-z_{i-1}, \; \;
\Delta z_{\text{tot},i} = \Delta z_{\text{up},i}+\Delta z_{\text{dn},i},
\end{align}

the discretization becomes

\begin{equation}
\begin{aligned}
\frac{2}{\Delta z_{\text{tot},i}\Delta z_{\text{dn},i}} \phi_{i-1}^{(n)}
- \frac{2}{\Delta z_{\text{up},i}\Delta z_{\text{dn},i}} \phi_{i}^{(n)}
+ \frac{2}{\Delta z_{\text{tot},i}\Delta z_{\text{up},i}} \phi_{i+1}^{(n)} 
& \\
= \; -\frac{\rho_i^{(n)}}{\epsilon_0} \; .
\end{aligned}
\end{equation}

The electric field is then approximated by the centered difference

\begin{equation}
E_i^{(n)} \approx - \frac{\phi_{i+1}^{(n)} - \phi_{i-1}^{(n)}}{z_{i+1}^{(n)}-z_{i-1}^{(n)}}.
\end{equation}

\section{Simulation}

For each GCM step ($\sim$15 min), charge density is computed from the triboelectric parameterization. 

Potentials and fields are then solved from the discretized Poisson system.

To avoid full particle tracking, charge in each grid cell is normalized by particle number density, carried forward, and rescaled at the next step. This preserves accumulated charging while keeping the scheme computationally tractable.

\section{Energy Density}
\label{app3}

Following electrostatic energy–density formulations developed for terrestrial thunderstorms \citep{Marshall2002JGRD..107.4052M, Coleman2003JGRD..108.4298C, Bruning2013JAtS...70.4012B},  the quasi-electrostatic expression for the local electrostatic energy density is:

\begin{equation}
  u \;=\; \tfrac{1}{2}\,\epsilon_0\,|E|^2 ,
  \label{eq:udens}
\end{equation}

where $\varepsilon_0$ is the permittivity of free space and $|E|$ is the electric–field magnitude in $V\,m^{-1}$.
In Mars’ $CO_2$ atmosphere, the relative permittivity is very close to unity, so $\varepsilon \approx \epsilon_0$.
Gas-discharge onset is governed by the reduced field $E/N$, where the neutral number density is obtained from the ideal gas law:

\begin{align}
    N = \frac{p}{k_{b}T} \;.
\end{align}

A spatially varying field limit is then built as

\begin{align}
    E_{lim} = \left( \frac{E}{N} \right)_{th} N \;,
\end{align}

where $\big(E/N\big)_{th}$ is a chosen threshold (e.g., Townsend/glow onset) for CO\(_2\).
A convenient anchoring uses a near–surface cap $E_{\mathrm{cap}}$ and a representative near–surface density $N_{\mathrm{ref}}$:

\begin{align}
    \left( \frac{E}{N} \right)_{th} = \frac{E_{cap}}{N_{ref}} \;. 
\end{align}

With this choice, $E_{\mathrm{lim}}\!\approx\!E_{\mathrm{cap}}$ near the surface and decreases aloft with $N$ (see Fig. \ref{fig:histogramm energy density2}).\\

The column (areal) excess energy is the vertical integral of the excess energy density:

\begin{align}
    U_{ex, areal} = \int \frac{1}{2} \epsilon_0 (E^2 - E_{lim}^2)\; dz \; .
\end{align}

Summing $U_{\mathrm{ex,areal}}$ over horizontal area yields the total available energy $U_{\mathrm{ex}}(t)$ for a region. \\

\FloatBarrier 
\clearpage

\end{appendix}
\end{document}